\documentclass[journal]{IEEEtran}

\usepackage{cite}

\usepackage{amsmath}
\usepackage{mathrsfs}

\usepackage{algorithmic}

\usepackage{array}

\ifCLASSOPTIONcompsoc
\usepackage[caption=false,font=normalsize,labelfont=sf,textfont=sf]{subfig}
\else
\usepackage[caption=false,font=footnotesize]{subfig}
\fi

\usepackage{url}
\usepackage{graphicx,amsmath,amssymb,amsfonts}
\usepackage{algorithmic,algorithm}

\usepackage{hyperref}
\hypersetup{colorlinks=true}

\usepackage{setspace}

\usepackage{xcolor}

\newtheorem{remark}{\textbf{Remark}}
\newtheorem{theorem}{\textbf{Theorem}}

\newtheorem{lemma}{\textbf{Lemma}}

\newtheorem{corollary}{\textbf{Corollary}}

\newtheorem{proposition}{\textbf{Proposition}}
\newtheorem{definition}{\textbf{Definition}}

\makeatletter
\newcommand{\rmnum}[1]{\romannumeral #1}
\newcommand{\Rmnum}[1]{\expandafter\@slowromancap\romannumeral #1@}
\makeatother

\usepackage{framed} 

\usepackage{cancel}

\begin{document}
	\bstctlcite{ref:BSTcontrol}

	\title{Integrating Over-the-Air Federated Learning and Non-Orthogonal Multiple Access: What Role can RIS Play?}

	\author{Wanli~Ni, \IEEEmembership{Graduate~Student~Member,~IEEE,}
				 Yuanwei~Liu, \IEEEmembership{Senior~Member,~IEEE,}
				 Zhaohui~Yang, \IEEEmembership{Member,~IEEE,}
			     Hui~Tian, \IEEEmembership{Senior~Member,~IEEE,}
				 and~Xuemin~Shen, \IEEEmembership{Fellow,~IEEE}
		\thanks{This work was supported by the National Key Research and Development Program of China (2018YFE0205502) and the BUPT Excellent Ph.D. Students Foundation. A short version of this paper has been presented at the IEEE INFOCOM Workshop on Distributed Machine Learning and Fog Networks, Virtual, May 2021 \cite{Ni2021Over}. \textit{(Corresponding author: Hui Tian.)}}
		\thanks{W. Ni and H. Tian are with the State Key Laboratory of Networking and Switching Technology, Beijing University of Posts and Telecommunications, Beijing 100876, China (e-mail: charleswall@bupt.edu.cn; tianhui@bupt.edu.cn).}
		\thanks{Y. Liu is with the School of Electronic Engineering and Computer Science, Queen Mary University of London, London E1 4NS, United Kingdom (e-mail: yuanwei.liu@qmul.ac.uk).}
		\thanks{Z. Yang is with the Department of Electronic and Electrical Engineering, University College London, London WC1E 7JE, United Kingdom (e-mail: zhaohui.yang@ucl.ac.uk).}
		\thanks{X. Shen is with the Department of Electrical and Computer Engineering, University of Waterloo, Waterloo, ON, N2L 3G1, Canada (e-mail: sshen@uwaterloo.ca).}
	}
	
\maketitle

\begin{abstract}
	With the aim of integrating over-the-air federated learning (AirFL) and non-orthogonal multiple access (NOMA) into an on-demand universal framework, this paper proposes a reconfigurable intelligent surface (RIS)-aided hybrid network by leveraging the RIS to flexibly adjust the decoding order of heterogeneous data.
	A new metric of computation rate is defined to measure the performance of AirFL users.
	Upon this, the objective of this work is to maximize the achievable hybrid rate by jointly optimizing the transmit power, controlling the receive scalar, and designing the reflection coefficients.
	Since the concurrent transmissions of all computation and communication signals are aided by the discrete phase-shifting elements at the RIS, the formulated problem (P0) is a challenging mixed-integer programming problem.
	To tackle this intractable issue, we decompose the original problem (P0) into a non-convex problem (P1) and a combinatorial problem (P2), which are characterized by the continuous and discrete variables, respectively.
	For the transceiver design problem (P1), the power allocation subproblem is first solved by difference-of-convex programming, and then the receive control subproblem is addressed by successive convex approximation, where the closed-form expressions of simplified cases are derived to obtain deep insights.
	For the reflection design problem (P2), a relaxation-then-quantization method is adopted to find a suboptimal solution for striking a trade-off between complexity and performance.
	Afterwards, an alternating optimization algorithm is developed to solve the non-linear non-convex problem (P0) iteratively.
	Finally, simulation results reveal that
	\rmnum{1}) the proposed RIS-aided hybrid network can support on-demand communication and computation efficiently,
	\rmnum{2}) the system performance can be improved by properly selecting the location of the RIS,
	and
	\rmnum{3}) the designed algorithms are also applicable to conventional networks with only AirFL or NOMA users.
\end{abstract}


\section{Introduction}
\IEEEPARstart{T}{he} prosperity of machine learning is expected to provide pervasive intelligence for beyond fifth generation (B5G) networks \cite{Chen2019Artificial, Liu2020When}, which will also make the ubiquitous data transmission more sensitive and demanding from privacy protection to resource utilization \cite{Lyu2019Optimal}.
By exploring the superposition property of wireless channels to compute a target function \cite{Nazer2007Computation} and allowing all participants to collaboratively train a shared model with keeping their raw data locally, over-the-air federated learning (AirFL) is greatly desired in many resource-limited and privacy-sensitive scenarios empowered by large-scale machine learning \cite{Yang2020TWC, Ni2021Federated}.
Additionally, as the number of connected devices and data traffic grow exponentially, the data-intensive applications would demand heterogeneous services such as high throughput, ultra reliability, and low latency \cite{Zhang20196G}.
Being helpful to overcome the challenges raised from the massive connectivity and resource scarceness, non-orthogonal multiple access (NOMA) has been deemed as one of the promising enablers for future networks \cite{Liu2017NOMA}.
To efficiently cope with the differentiated requirements of the hybrid networks consisting of both AirFL and NOMA users, it is particularly important to design an on-demand universal framework that is capable of integrating computation and communication together for meeting the diversity needs of the forthcoming flourish applications.

Since the conflicts between the spectrum scarcity of existing networks and the bandwidth requirements of emerging applications are ever-increasingly striking \cite{Zhang20196G}, the co-design of distributed machine learning and wireless communication is essential for B5G networks to bring considerable gains in terms of latency reduction \cite{Zhu2020Broadband}, spectrum efficiency \cite{Qi2020Integrated}, energy consumption \cite{Yang2021Energy}, and hardware cost.
To this end, understanding the similarity and difference between AirFL and NOMA is extremely valuable for the seamless integration of them.
On the one hand, the uplink transmissions of AirFL are over the same multiple-access channel \cite{Zhu2020Broadband}, which is similar to that of NOMA technique \cite{Liu2017NOMA}.
On the other hand, the AirFL concentrates on the computation result rather than the individual data itself \cite{Ni2021Federated, Qi2020Integrated}, which is different from NOMA.
Exploiting the mentioned similarity, the signals from hybrid users can be multiplexed on the same channel simultaneously.
However, it is still not a trivial task to provide heterogeneous services over limited radio spectrum, due to the processing difficulties of the mixed signals.
Therefore, it is highly desired for the hybrid network to deploy the reconfigurable intelligent surface (RIS) for dynamically tuning the wireless channels, which plays a key role to flexibly adjust the signal processing order of hybrid users.

\subsection{Related Works}
Recently, both AirFL and NOMA have attracted compelling attention, but most previous studies have focused on the separate implementation of AirFL and NOMA in diverse scenarios such as the works in \cite{Yang2020TWC, Amiri2020Machine, Zhu2021One, Ni2021Federated, Liu2017NOMA, Liu2017Grouping, Zhao2017Spectrum, Cui2018QoE}.
However, the network architecture may be isolated and dedicated if designed independently, which inevitably makes the flexibility and versatility of existing networks insufficient.
By establishing an explicit expression for the packet error rate and the convergence rate of federated learning, Chen \textit{et. al} \cite{Chen2020A} provided a joint learning and communication algorithm to reduce the training loss.
Different from the conventional communication paradigms based on orthogonal multiple access, by exploiting the superposition property of wireless channels, multiple users are able to implement model aggregation (i.e., function computation) over-the-air with the help of the remote base station (BS) \cite{Nazer2007Computation, Yang2020TWC, Ni2021Federated}.
Thus, the new technique called AirFL is recently developed, which is beneficial to alleviate the communication latency dramatically \cite{Zhu2020Broadband, Zhu2021One, ChenQW18WCL}.
In order to improve the convergence rate and prediction accuracy of AirFL, the authors of \cite{Yang2020TWC} proposed a difference-of-convex (DC) programming based method to optimize the device selection and active beamforming for aggregation error minimization.
Additionally, to overcome the unfavorable noise brought by the wireless environment, Jiang \textit{et. al} \cite{Jiang2019Over} deployed the RIS to alleviate the signal distortion of AirFL.
Moreover, our previous work in \cite{Ni2021Federated} jointly designed the transceiver and reflection settings to further enhance the learning performance of AirFL with the aid of multiple intelligent surfaces.

Due to the interference caused by the spectrum sharing among multiple users, efficient resource management and interference mitigation are the fundamental research challenges for NOMA networks \cite{Liu2017NOMA, Liu2017Grouping, Zhao2017Spectrum, Cui2018QoE}.
To improve the spectrum efficiency, Zhao \textit{et. al} \cite{Zhao2017Spectrum} formulated a joint problem of spectrum allocation and power control in NOMA enhanced heterogeneous networks.
Also, by adopting the matching theory and optimization methods, the authors of \cite{Cui2018QoE} aimed to provide better user experience in the multi-cell NOMA networks with multiple subcarriers.
Besides, considering the problems in RIS-aided NOMA networks, cases from ideal and non-ideal RIS to single and multi-cell were discussed in \cite{Mu2020Exploiting} and \cite{Ni2021Resource}, respectively.
Motivated by the ability to learn from long-term evolution, Liu \textit{et. al} \cite{Liu2020RIS} developed reinforcement learning based approaches to solve the RIS deployment problem in a highly dynamic NOMA network.
However, most prior works ignore the system heterogeneity from device status to service requirements.
For instance, the computation units and energy supplies of many Internet-of-Things (IoT) devices are usually insufficient, which makes them prefer to data offloading rather than local computing.
Moreover, future networks may involve a multitude of user devices in most cases, their services and applications can be quite various from one another \cite{Yonina2021Federated}.
Therefore, these behavior heterogeneities should be carefully addressed to fulfill the intelligence vision of next-generation wireless networks.

\subsection{Motivations and Challenges}
Again, although previous studies have well investigated the separate design, the network-wide convergence of distributed learning and wireless communication is still a nascent field and many open issues deserve exploration.
The goal of this work is to consider:
Is it possible to seamlessly merge distributed machine learning into traditional wireless networks with constrained resources, and what role can RIS play in it?
Note that the integration of AirFL and NOMA makes the received superposition signal at the BS quite complicated, namely two different types of users are multiplexed on the same channel, which also makes the critical issues of integrated networks different from that of conventional networks consisting of only AirFL or NOMA users.
In summary, this entails the following challenges.
\begin{itemize}
	\item 
	At present, it is generally difficult for the computation-centric AirFL to be integrated into the communication-centric wireless networks.
	So far a huge challenge is how to distinguish the homogeneous signals received simultaneously from different types of users, which motivates us to rethink existing techniques as reliable solutions for building a universal framework.
	\item
	To date, there is a lack of rate-like criterion to evaluate the aggregation performance for the collaborative learning of AirFL users.
	Most existing works address the computation error issue only by minimizing the signal distortion measured by the mean-square-error (MSE) but neglect the performance comparison requirements of different user types, which thus motivates us to give a novel metric for defining a comparable evaluation methodology.
	\item
	In addition, due to the practical implementation concerns, the discrete phase shifts of the RIS make the joint optimization problem be a mixed-integer programming, which is highly challenging.
	Thus, the resource allocation problem in terms of transceiver control and reflection design is NP-hard and is non-trivial to solve.
	Not to mention that the previous researches on RIS-aided hybrid networks are insufficient, or even almost none.
\end{itemize}

\subsection{Contributions and Organization}
To overcome the aforementioned challenges, we first combine the communication-centric NOMA users and the computation-centric AirFL users into one concurrent uplink transmission.
Then, the RIS is leveraged to adjust the signal processing order of successive interference cancellation (SIC) decoding via tuning the channel quality of hybrid users on demand.
By doing so, the concurrent transmission of heterogeneous data is capable of providing the high spectrum efficiency and low latency services to satisfy diverse requirements.
To the best of our knowledge, this is the first effort to provide a universal framework by leveraging the RIS to unify NOMA communication and collaborative computation via concurrent uplink transmission.
The main contributions of this paper can be summarized as follows:
\begin{enumerate}
	\item
	We integrate AirFL and NOMA in a universal framework by combining computation and communication signals into one concurrent transmission, in which the important role of RIS is revealed.
	In order to unify the performance metrics of different user types, we define an exact expression of computation rate to evaluate the learning behavior of AirFL users.
	Based on which, we aim to maximize the hybrid rate (defined as the weighted sum of the communication rate and the computation rate) by jointly optimizing the continuous transceiver scalars and the discrete phase shifts at the RIS, while satisfying the minimum rate requirements of NOMA users and the maximum aggregation error of AirFL users.
	The formulated resource allocation problem is analyzed to be a mixed-integer non-linear programming (MINLP), where the network characterization is represented effectively.
	\item 
	In an effort to solve the transceiver design problem with continuous variables, we first use DC programming and successive convex approximation (SCA) methods to transform decoupled non-convex subproblems into convex ones.
	Then, in order to tackle the discrete variables in the reflection design problem, we adopt linear substitution and semidefinite relaxation (SDR) methods to handle the combinatorial optimization challenge.
	Furthermore, regarding the simplified cases without quality-of-service (QoS) and MSE constraints, we provide explicit closed-form solutions for power allocation at the NOMA users and the receive control at the BS, respectively.
	After that, we develop an alternating optimization algorithm to solve the intractable original problem with low complexity.
	\item 
	We conduct extensive numerical simulations to demonstrate the effectiveness of our proposed three-step alternating optimization algorithm for the designed RIS-aided hybrid network.
	Specifically, benchmark schemes from ideal phase shifts to non-ideal cases are compared to evaluate the performance gain under different conditions.
	We also show that the RIS configurations such as the number of passive elements, the quantization resolution and the deployment location have significant impacts on the achievable communication-computation hybrid rate.
	Therefore, these aforementioned factors should be carefully optimized in practice to strike a good balance between hardware cost and network performance.
\end{enumerate}

The rest of this paper is organized as follows.
First, system model of RIS-aided hybrid networks is given in Section \ref{section_system_model}.
Then, a resource allocation problem is formulated in Section \ref{section_problem_formulation}.
Next, an alternating optimization algorithm is proposed in Section \ref{alternating_optimization}, where the corresponding convergence and complexity are analyzed.
Finally, simulations results are presented in Section \ref{section_simulation}.
The conclusion is drawn in Section \ref{section_conclusion}, where the role of RIS is summarized.

\begin{figure} [t!]
	\centering
	\includegraphics[width=3.5 in]{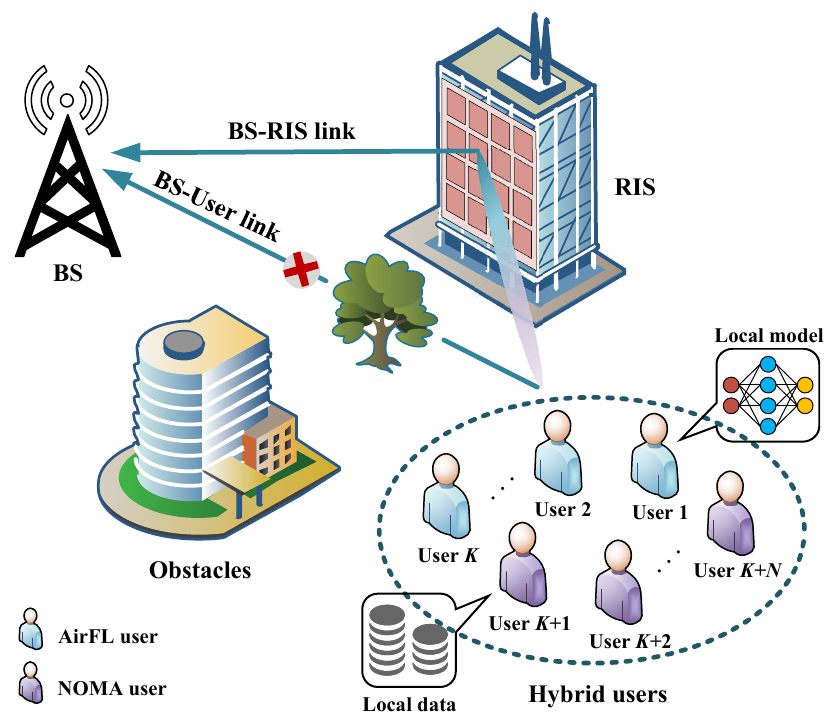}
	\caption{An illustration of the RIS-aided network, where one RIS equipped with $M$ reflecting elements is deployed to assist the communication and computation in concurrent transmissions from the users to the BS. Specifically, there are two types of users, namely the NOMA users and the AirFL users, are served by the BS simultaneously. Meanwhile, the direct links between the users and the BS are not available due to the blockage.}
	\label{system_model}
\end{figure}

\section{System Model} \label{section_system_model}
\subsection{Network and Channel Model}
As illustrated in Fig. \ref{system_model}, we consider a RIS-aided wireless network consisting of one single-antenna BS, $I$ single-antenna hybrid users, and one RIS equipped with $M$ reflecting elements.
Note that the hybrid users are composed of two types, i.e., the NOMA users and the AirFL users.
Both of them are served by the non-orthogonal communication over the wireless multiple-access channels.
The set of hybrid users is indexed by $\mathcal{I} = \{ 1,2, \ldots, K, K+1, K+2, \ldots, K+N \}$, where the sets of AirFL users and NOMA users are indexed by $\mathcal{K} = \{ 1,2, \ldots, K\}$ and $\mathcal{N} = \{ K+1, K+2, \ldots, K+N\}$, respectively.
The set of reflecting elements is denoted by $\mathcal{M} = \{1, 2, \ldots, M\}$.
The $M \times M$ diagonal reflection matrix of the RIS is denoted by $\mathbf{\Theta} = {\rm diag} ( e^{j\theta_1}, e^{j\theta_2}, \dots, e^{j\theta_M} )$, where $\theta_{m}$ represents the phase shift of the $m$-th reflecting element.

In practice, the phase shifts of the RIS are limited to a finite number of discrete values due to the inevitable hardware impairments brought by the physical implementation \cite{Di2020Hybrid, Wu2020Beamforming, Mu2020Exploiting, Zhang2020IRS}.
For this reason, we assume that the resolution of phase shifts is encoded by $b \in \mathbb{N}^{+}$ bits, and each reflecting element has $\bar{B} = 2^{b}$ possible configuration patterns.
By using the common uniform quantization method, the feasible set of the discrete phase shifts can be expressed as
\begin{equation}
\mathcal{A} = \left\lbrace \frac{1}{2} \Delta, \ \frac{3}{2} \Delta, \ \frac{5}{2} \Delta, \ \ldots, \ \frac{2\bar{B}-1}{2} \Delta \right\rbrace,
\end{equation}
where $\Delta = 2 \pi / \bar{B}$ represents the phase resolution.
Then, one can know that the RIS with $M$ reflecting elements can be configured to one of $\bar{B}^{M}$ operation modes \cite{Wu2020Beamforming}.
Furthermore, we assume that the direct links between the users and the BS are blocked by unfavorable propagation conditions such as obstacles \cite{Huang2019EE, Zhang2020IRS, Mu2020Joint, Huang2020Reconfigurable}.
Due to the light-of-sight (LoS) component provided by the RIS configured at a high building, we use Rician model to formulate all RIS-related channels \cite{Zhang2020IRS, Mu2020Joint}.
Specifically, the small-scale fading channels of the BS-RIS link and the user-RIS link are denoted by $\boldsymbol{g} \in \mathbb{C}^{M \times 1}$ and $\boldsymbol{h}_{i} \in \mathbb{C}^{M \times 1}, \forall i \in \mathcal{I}$, respectively.

When the RIS is deployed on the facade of a building, the distance from the RIS to the BS and hybrid users can be given as $d_{i}, \forall i \in \{0\} \cup \mathcal{I}$, where $i = 0$ for the BS, while $i \in \mathcal{I}$ for the hybrid user.
In terms of the large-scale fading, the distance-dependent path loss model is adopted as $L_{i} = \varsigma_0 (d_{i})^{-\alpha}, \forall i \in \{0\} \cup \mathcal{I} $, where $\varsigma_0$ is the path loss at the reference distance of 1 meter and $\alpha \ge 2$ is the path loss exponent.
Therefore, the combined channel coefficient from the $i$-th user to the BS via the RIS can be given by
\begin{equation} \label{combined_channel}
\bar{h}_{i} = \sqrt{L_{0}L_{i}} \boldsymbol{g}^{\rm H} \mathbf{\Theta} \boldsymbol{h}_{i} = \varsigma_0 \sqrt{(d_{0} d_{i})^{-\alpha}}  \boldsymbol{v}^{\rm H} \boldsymbol{\Phi}_{i}, \ \forall i \in \mathcal{I},
\end{equation}
where $\boldsymbol{\Phi}_{i} = {\rm diag} ( \boldsymbol{g}^{\rm H} ) \boldsymbol{h}_{i}$ and $\boldsymbol{v} = \left[ v_{1}, v_{2}, \dots, v_{M} \right] ^{\rm H}$ with $v_{m} = e^{j\theta_{m}}$.
Note that the vector $\boldsymbol{v}$ is able to well contain all information of the phase shifts of the RIS by separating the vector $\boldsymbol{v}$ from the reflective channels $\boldsymbol{\Phi}_{i}$, which is significantly beneficial to our derivations in Section \ref{phase_shifts_design}.
Moreover, we assume that the perfect channel state information (CSI) of all channels is available at the BS via the estimation methods described in \cite{Wang2019Channel} and \cite{Liu2020Matrix}.
Meanwhile, all channels are assumed to obey the frequency-flat and quasi-static fading model \cite{Huang2020Reconfigurable, Ni2021Resource}.

\begin{figure*} [t!]
	\centering
	\includegraphics[width=7 in]{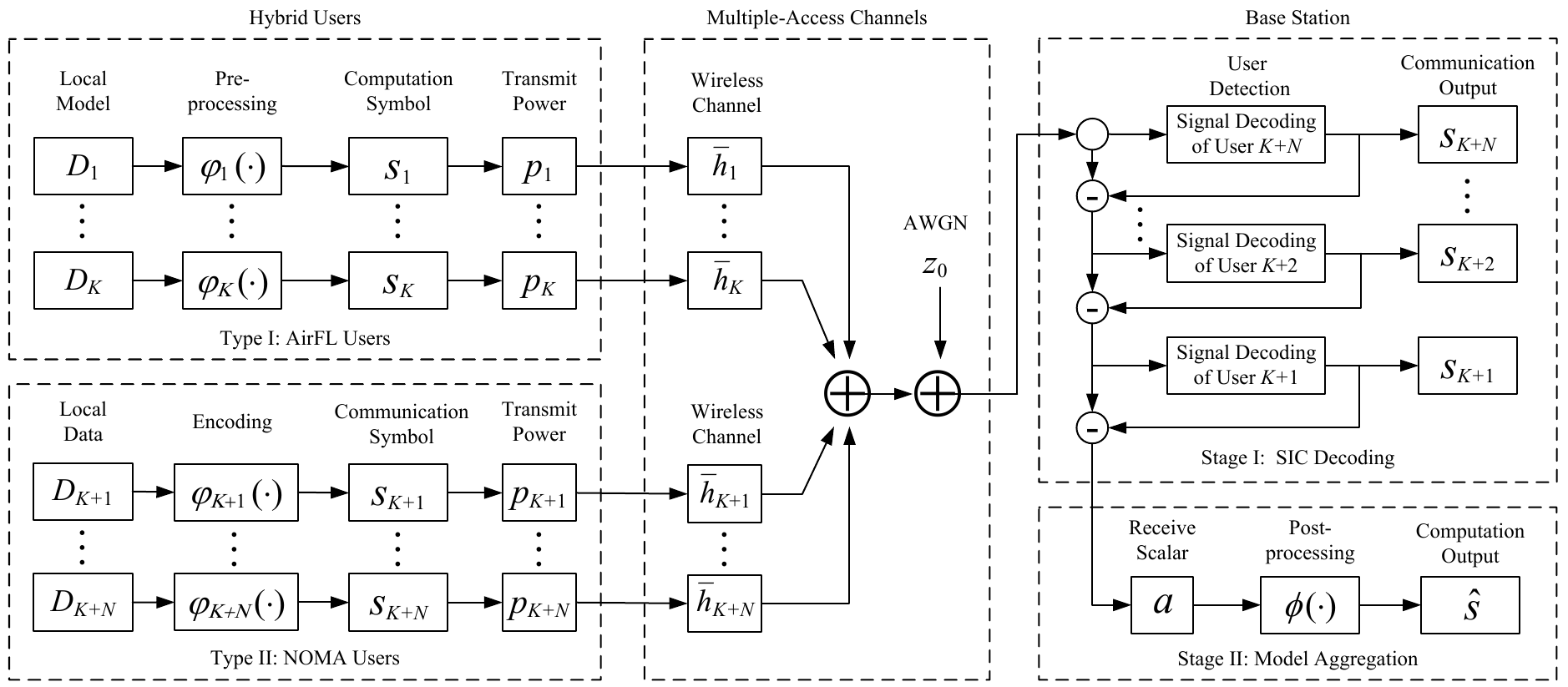}
	\caption{Block diagram of the RIS-aided hybrid network consisting of both AirFL and NOMA users.}
	\label{block_diagram}
\end{figure*}

\subsection{RIS-Aided Concurrent Transmission}
By serving both the AirFL and NOMA users on the same time-frequency resource, the superposed signal received at the BS is expressed as $y = \sum \nolimits_{i=1}^{K+N} \bar{h}_{i} p_{i} s_{i} + z_{0}$, expanded as
\begin{equation} \label{concurrent_transmission}
y = \underbrace{\sum \nolimits_{k=1}^{K} \bar{h}_{k} p_{k} s_{k}}_{\mathbf {AirFL~users}}
   + \underbrace{\sum \nolimits_{n=K+1}^{K+N} \bar{h}_{n} p_{n} s_{n}}_{\mathbf {NOMA~users}}
   + \underbrace{z_{0}}_{\mathbf {noise}},
\end{equation}
where $p_{k}$ ($p_{n}$) is the power scalar at the $k$-th ($n$-th) user,
$s_{k}$ is the transmit signal (i.e., computation symbol) of local model after pre-processing at the $k$-th AirFL user, and $s_{n}$ is the information-bearing signal (i.e., communication symbol) of the $n$-th NOMA user.
$z_{0} \sim \mathcal{CN}(0, \sigma^{2})$ is the additive white Gaussian noise (AWGN) with zero mean, and $\sigma^{2}$ is the noise power at the BS.
The reason for distinguishing signals from different users in (\ref{concurrent_transmission}) is provided in the remark below.

\begin{remark}
	\emph{Note that the transmission goals of the NOMA users and the AirFL users are very different in nature.
		To be specific, the NOMA users are so \textit{communication-centric} that they expect to maximize individual data rate and/or their sum rate \cite{Liu2017Grouping, Liu2017NOMA, Zuo2020Resource}.
		On the contrary, the AirFL users are \textit{computation-centric} such that all of them anticipate minimizing the global aggregation error to improve the learning performance \cite{Yang2020TWC, Zhu2020Broadband, Ni2021Federated}.
		As a result, compared with the NOMA users, these AirFL users concentrate more on the computation results of the transmitted data rather than the individual data itself.}
\end{remark}

To facilitate the optimization of transmit power, the information symbols of all users are assumed to be statistically independent and are normalized with unit variance, i.e., $\mathbb{E}(|s_i|^2)=1$ and $\mathbb{E}(s_i^{\rm H} s_j)=0, \forall i \ne j$.
Then, the transmit power constraint at the $i$-th user can be given by
\begin{equation}\label{power_constraint}
\mathbb{E}(| {p_{i}} s_{i} |^2) = \left| p_{i} \right| ^{2} \le P_{i}, \ \forall i \in \mathcal{I},
\end{equation}
where $P_{i} > 0$ is the maximum transmit power at the $i$-th user.

Since all users are served by the entire bandwidth simultaneously, interference management is an essential issue for the hybrid network composed of both AirFL and NOMA users, especially the interference from inter-type users.
Using SIC techniques, the BS can first decode the signals of strong users in a successive order to remove the interference, and then it is able to obtain the superposed signal of weak users \cite{Liu2017NOMA}.
To take advantage of this technique, all NOMA users are expected to become strong users and all AirFL users are anticipated to be treated as weak users in the hybrid network.
By doing so, the BS is capable of eliminating the interference from NOMA users for these AirFL users via SIC decoding before executing global model aggregation.

Sparked by the above innovative idea, the block diagram of the RIS-aided hybrid network is illustrated in Fig. \ref{block_diagram}.
From this figure, one can notice that it is highly desired for the AirFL and NOMA hybrid network to flexibly adjust the signal processing order of hybrid users by effectively deploying the RIS to change multiple-access channels.
In order to guide the adjustment of SIC decoding order (i.e., tuning the AirFL users into weak users), the following constraint should be met by judiciously optimizing the phase shifts at the RIS, i.e.,
\begin{equation} \label{state_factor_order}
\left| \bar{h}_{k} \right| ^{2} \le \left| \bar{h}_{K+1} \right| ^{2} \le \left| \bar{h}_{K+2} \right| ^{2} \le \cdots \le \left| \bar{h}_{K+N} \right| ^{2}, \ \forall k \in \mathcal{K}.
\end{equation}
In this paper, the SIC decoding order only depends on the uplink channel gains of hybrid users, and is assumed to be fixed for simplicity.
The design of more complicated SIC decoding schemes may further improve the attainable system performance, but this is beyond the scope of this work.
Note that the RIS in this paper acts as a coordinator to dynamically adjust the channel coefficient between two types of users, which makes the SIC decoding order more flexible.
Compared with many previous studies that only use RIS to improve the performance of existing networks, the RIS in our proposed universal framework plays a very important role in maintaining the decoding order given in (\ref{state_factor_order}).

\subsection{Achievable Hybrid Rate}
Due to the co-existence of two types of users, the performance metrics of the considered network are composed of two parts: communication rate for NOMA users and computation rate for AirFL users, the details are given below.
\subsubsection{NOMA Communication Rate}
According to the fixed SIC decoding order given in (\ref{state_factor_order}), the BS first successfully decodes the signal of the strongest user, and then it successively removes the interference to decode the desired signal of the second strongest user, and so on.
As such, the SINR of the $n$-th NOMA user at the BS can be written as
\begin{equation}
\gamma_{n} = \frac{ \left| p_{n} \right| ^{2} \left| \bar{h}_{n} \right| ^{2} }{ \sum_{k=1}^{K} \left| p_{k} \right| ^{2} \left| \bar{h}_{k} \right| ^{2} + \sum_{i=K+1}^{n-1} \left| p_{i} \right| ^{2} \left| \bar{h}_{i} \right| ^{2} + \sigma^{2}}, \ \forall n \in \mathcal{N}.
\end{equation}

It is worth pointing out that the computation symbols from the weak AirFL users are not removed and are treated as interference by the strong NOMA users.
Based on the Shannon capacity theory, the achievable uplink data rate of the $n$-th NOMA user can be given by
\begin{equation} \label{NOMA_rate}
R_{n} = B \log_{2} \left( 1 + \frac{ \left| p_{n} \right| ^{2} \left| \bar{h}_{n} \right| ^{2} }{ \sum_{i=1}^{n-1} \left| p_{i} \right| ^{2} \left| \bar{h}_{i} \right| ^{2} + \sigma^{2}} \right), \ \forall n \in \mathcal{N},
\end{equation}
where $B$ is the available bandwidth at the BS.
Hence, the uplink sum rate of these NOMA users can be given as $R_{\rm NOMA} = \sum_{n=K+1}^{K+N} R_{n}$.

\subsubsection{AirFL Computation Rate}
When all signals from NOMA users have been decoded via SIC technique, the BS can subtract them from the received superposition signal to obtain the aggregated signal of all AirFL users for model aggregation, which can be written as
\begin{equation}
\hat{y} = \sum \nolimits_{k=1}^{K} \bar{h}_{k} p_{k} s_{k} + z_{0}.
\end{equation}

Next, as shown in Fig. \ref{block_diagram}, the BS applies a receive scalar $a \in \mathbb{C}$ to the aggregated signal $\hat{y}$, then the reconstructed signal of interest to AirFL users can be given by
\begin{equation}
	\hat{s} = \frac{a}{K} \hat{y} = \frac{a}{K} \sum \nolimits_{k=1}^{K} \bar{h}_{k} p_{k} s_{k} + \frac{a}{K} z_{0},
\end{equation}
where the post-processing function $1/K$ is a model aggregation operation using arithmetic mean that can be replaced with other nomographic functions summarized in \cite{ChenQW18WCL, Zhu2019AirComp, Qi2020Integrated}.

Compared with the desired computation result from all AirFL users, denoted by
\begin{equation}
	s = \frac{1}{K} \sum \nolimits_{k=1}^{K} s_{k},
\end{equation}
the computation distortion of $\hat{s}$ with respect to $s$ is defined as ${\rm MSE} (\hat{s}, s) \triangleq \mathbb{E} ( |\hat{s} - s|^2)$ \cite{Yang2020TWC}, i.e.,
\begin{equation} \label{MSE_s}
{\rm MSE} \left( \hat{s}, s \right)
= \frac{1}{K^{2}} \sum \nolimits_{k=1}^{K} \left| a \bar{h}_{k} p_{k} - 1 \right| ^2 + \frac{ |a|^{2} \sigma^2}{K^{2}},
\end{equation}
where ${\rm MSE} (\hat{s}, s)$ can be regarded as the interference-plus-noise power experienced by AirFL users.
Since ${\rm MSE} (\hat{s}, s)$ can effectively measure the computation distortion of all AirFL users, we are now interested in quantifying the system rate for the collaborative computation.
To this end, the achievable computation rate of global model aggregation is defined as follows.

\begin{definition} \label{definition_AirFL_rate}
	\emph{For AirFL users, the computation rate of model aggregation over a noisy multiple-access channel can be defined as \cite{Ni2021Over, Hu2021ICASSP}
	\begin{equation} \label{AirFL_rate}
		R_{\rm AirFL} = B \log_{2} \left( 1 + \frac{ \mathbb{E} ( |\hat{s}|^2) - {\rm MSE} (\hat{s}, s) }{ {\rm MSE} (\hat{s}, s)} \right),
	\end{equation}
	where $\mathbb{E} ( |\hat{s}|^2)$ can be deemed as the total received power of the reconstructed signal.}
\end{definition}

Note that the computation rate given in Definition \ref{definition_AirFL_rate} is characterized by the similar logarithm form of the communication rate in (\ref{NOMA_rate}), which unifies the performance metrics of two different types of users and also makes them comparable.
Therefore, based on the derived rate expressions in (\ref{NOMA_rate}) and (\ref{AirFL_rate}), the achievable hybrid rate of the considered network can be written as
\begin{equation} \label{hybrid_rate}
R_{\rm Hybrid} = (1 - \lambda) R_{\rm NOMA} + \lambda R_{\rm AirFL},
\end{equation}
where $\lambda \in [0,1]$ is a weight parameter for the performance trade-off between the NOMA users and the AirFL users.
An exact expression of the achievable hybrid rate is given in (\ref{exact_hybrid_rate}).
Meanwhile, it can be noted that the choice of this weight parameter has a great impact on the resource allocation between the two types of users.

\begin{figure*}[t]
	\begin{align} \label{exact_hybrid_rate}
	R_{\rm Hybrid}
	= \left( 1 \!-\! \lambda \right) \underbrace{ \sum \limits_{ n \in \mathcal{N} } B\!\log_{2} \left( 1 \!+\! \frac{ \left| p_{n} \right| ^{2} \left| \bar{h}_{n} \right| ^{2} }{ \sum \limits_{i=1}^{n-1} \left| p_{i} \right| ^{2} \left| \bar{h}_{i} \right| ^{2} + \sigma^{2}} \right) }_{\mathbf {Communication~rate}}
	\!+\! \lambda\!\underbrace{ B\!\log_{2} \left( \frac{ \sum \limits_{ k \in \mathcal{K} } \left| a \bar{h}_{k} p_{k} \right| ^2 + |a|^{2} \sigma^2 }{ \sum \limits_{ k \in \mathcal{K} } \left| a \bar{h}_{k} p_{k} - 1 \right| ^2 + |a|^{2} \sigma^2 } \right)}_{\mathbf {Computation~rate}}
	\end{align}
	\hrulefill
\end{figure*}

\section{Problem Formulation and Decomposition} \label{section_problem_formulation}
Given the considered system model of RIS-aided hybrid networks, the objective of this paper is to maximize the achievable hybrid rate by jointly designing the transmit power $\boldsymbol{p}$ at the users, controlling the receive scalar $a$ at the BS, and tuning the phase shifts $\boldsymbol{v}$ at the RIS.
Subject to the QoS requirement, the aggregation error demand, and the maximum power constraint, the hybrid rate maximization problem can be formulated as
{\allowdisplaybreaks[4]
\begin{subequations} \label{original_problem}
	\begin{align}
		(\mathcal{P}0): 
		\label{original_objective}
		\max \limits_{\boldsymbol{p}, a, \boldsymbol{v}}
		\ & (1 - \lambda) R_{\rm NOMA} + \lambda R_{\rm AirFL} \\
		{\rm s.t.}
		\label{original_constraint_QoS}
		\ & R_{n} \ge R_{n}^{\min}, \ \forall n \in \mathcal{N}, \\
		\label{original_constraint_MSE}
		\ &{\rm MSE} \left( \hat{s}, s \right) \le \varepsilon_0, \\
		\label{original_constraint_phase_shift}
		\ & \theta_{m} \in \mathcal{A}, \ \forall m \in \mathcal{M}, \\
		\label{constraints_of_power_and_channel}
		\ & {\rm (\ref{power_constraint}) \ and \ (\ref{state_factor_order})},
	\end{align}
\end{subequations}
where
$\boldsymbol{p} = [p_{1}, p_{2}, \ldots, p_{K+N}]^{\rm T}$ is the transmit power vector,
$R_{n}^{\min}$ is the minimum communication rate required by the $n$-th NOMA user,
$\varepsilon_0 > 0$ is the maximum computation error allowed by AirFL users.}
Specifically,
the objective function is given in (\ref{original_objective}).
The QoS requirement is guaranteed by constraint (\ref{original_constraint_QoS}).
The MSE tolerance of global aggregation is represented in constraint (\ref{original_constraint_MSE}).
The discrete phase shift is drawn in constraint (\ref{original_constraint_phase_shift}).
Constraints (\ref{power_constraint}) and (\ref{state_factor_order}) in (\ref{constraints_of_power_and_channel}) are related to transmit power and SIC decoding, respectively.
Due to the coupling of multiple continuous and discrete variables in both the objective function and constraints, the formulated problem (\ref{original_problem}) is intractable.

The formulated problem (\ref{original_problem}) with both continuous and discrete variables is a MINLP problem and it generally belongs to NP-hard, which is difficult to solve optimally. Furthermore, the original problem (\ref{original_problem}) is still very hard to tackle efficiently even for the case with only one user type, i.e., $\lambda = 0$ or $\lambda = 1$, due to the non-convex objective function and constraints from (\ref{original_objective}) to (\ref{original_constraint_MSE}) as well as the combinatorial features of discrete phase shifts in (\ref{original_constraint_phase_shift}).
On the whole, there is no standard optimization approach to find the global optimal solution for the NP-hard problem (\ref{original_problem}) directly.
To address this MINLP problem effectively for obtaining high quality suboptimal solutions, we propose to decouple it into the following two subproblems:

\begin{enumerate}
	\item \emph{Transceiver Design}: Given the discrete phase shifts of the RIS, our first subproblem is to maximize the hybrid rate by dynamically controlling the transmit power at users and optimizing the receive scalar at the BS.
	As a result, subject to the QoS requirements of NOMA users, the MSE tolerance of AirFL users, and the transmit power constraints, the non-convex and non-linear problem for transceiver design can be given by
	\begin{subequations} \label{transceiver_problem}
		\begin{align}
		(\mathcal{P}1): 
		\label{transceiver_problem_objective}
		\max \limits_{ \boldsymbol{p}, a }
		\ & (1 - \lambda) R_{\rm NOMA} + \lambda R_{\rm AirFL} \\
		{\rm s.t.}
		\ & {\rm (\ref{power_constraint}), (\ref{original_constraint_QoS}) \ and \ (\ref{original_constraint_MSE})}.
		\end{align}
	\end{subequations}
	\item \emph{Reflection Design}: Given the transmit power and the receive scalar, our second subproblem is to improve the achievable performance by judiciously tuning the discrete phase shifts of the RIS.
	Therefore, subject to the channel quality order, data rate and error-related constraints for users, as well as the discrete phase shifts constraints, the combinatorial optimization problem for reflection design can be written as
	\begin{subequations} \label{RIS_problem}
		\begin{align}
		(\mathcal{P}2): 
		\label{RIS_problem_objective}
		\max \limits_{ \boldsymbol{v} }
		\ & (1 - \lambda) R_{\rm NOMA} + \lambda R_{\rm AirFL} \\
		{\rm s.t.}
		\ & {\rm (\ref{state_factor_order}), (\ref{original_constraint_QoS}), (\ref{original_constraint_MSE}) \ and \ (\ref{original_constraint_phase_shift})}.
		\end{align}
	\end{subequations}
\end{enumerate}

Note that the transceiver design subproblem (\ref{transceiver_problem}) is very challenging even when only the transmit power subproblem or the receive scalar subproblem is considered.
Moreover, the reflection design subproblem (\ref{RIS_problem}) is still intractable even we relax the phase shifts into continuous ones.
Although subproblems (\ref{transceiver_problem}) looks simpler than the original problem (\ref{original_problem}), it is still almost impossible to derive closed-form solutions for the transmit power and receive scalar in the considered RIS-aided hybrid network.
Not to mention that the combinatorial feature of subproblem (\ref{RIS_problem}) makes it NP-hard and much more complicated, e.g., there are $B^{M}$ operation modes of the phase shifts to be tested if the exhaustive search is used \cite{Wu2020Beamforming, Di2020Hybrid}.
To avoid an unacceptable computation complexity, one common approach is to transform these non-linear and non-convex subproblems of transceiver control and reflection design into convex ones, and then standard methods can be invoked to solve them separately and alternatively over iterations.
As a consequence, the alternating optimization technique can be adopted as an intuitive approach to solve the MINLP problem (\ref{original_problem}) in an efficient manner.

\begin{figure*}[t]
	\centering
	\includegraphics[width=7 in]{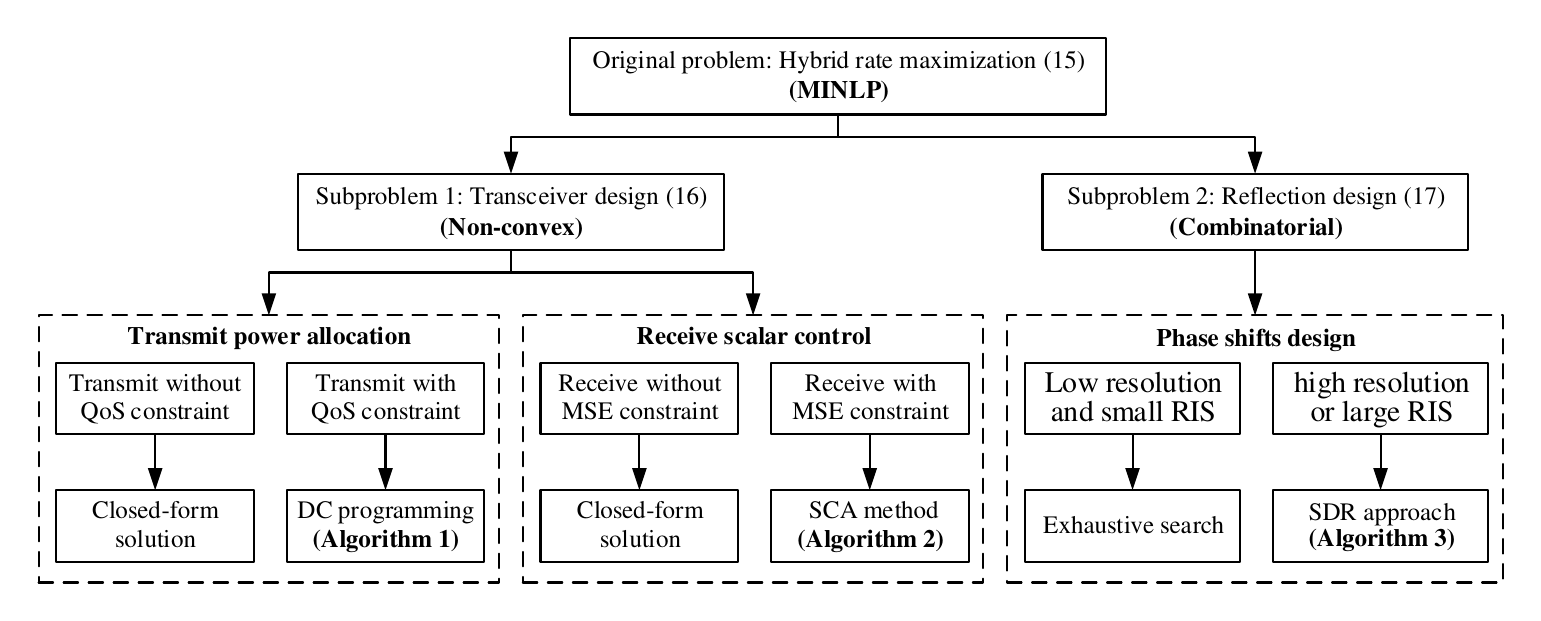}
	\caption{An overview of the problem decomposition and proposed methods to subproblems.}
	\label{problem_roadmap}
\end{figure*}

For providing a clear illustration, Fig. \ref{problem_roadmap} gives an overview of the roadmap to solve the hybrid rate maximization problem, especially the connections between key optimization subproblems and the designed algorithms.
Firstly, the original MINLP problem (\ref{original_problem}) is decomposed into two subproblems: the non-convex transceiver design subproblem (\ref{transceiver_problem}) and the combinatorial reflection design problem (\ref{RIS_problem}), which are characterized by the continuous variables and discrete variables.
Secondly, the subproblems of transmit power allocation, receive scalar control, phase shifts design are solved sequentially in Section \ref{alternating_optimization} by using common optimization methods such as DC programming, SCA, and SDR.
Notably, closed-form solutions for some simplified cases are derived to provide useful insights.

%


\section{Alternating Optimization} \label{alternating_optimization}

\subsection{Transmit Power Allocation}
With given receive scalar $a$ in subproblem (\ref{transceiver_problem}), we focus on optimizing the transmit power $\boldsymbol{p}$ in this subsection.
Before solving this problem, we notice that the power allocation strategies of AirFL and NOMA users, i.e., $\boldsymbol{p}_{\rm A} = [p_{1}, p_{2}, \ldots, p_{K}]^{\rm T}$ and $\boldsymbol{p}_{\rm N} = [p_{K+1}, p_{K+2}, \ldots, p_{K+N}]^{\rm T}$, are coupled in the objective function (\ref{transceiver_problem_objective}), but separated in the constraints.
Meanwhile, considering that the signals of NOMA users are processed first at the BS, followed by the signals of AirFL users, which has been illustrated in Fig. \ref{block_diagram}.
Thus, we propose to first fix the transmit power $\boldsymbol{p}_{\rm A}$ of AirFL users, and tackle the power allocation problem of NOMA users, then repeat this in turn until convergence.
The details are given below.

\subsubsection{Power allocation of NOMA users}
By following the above ideas, given $\boldsymbol{p}_{\rm A}$, the power allocation problem of NOMA users can be equivalently written as
\begin{subequations} \label{transmit_problem_NOMA}
	\begin{align}
	\label{transmit_problem_NOMA_objective}
	\max \limits_{ \boldsymbol{p}_{\rm N}}
	\ & \sum_{n \in \mathcal{N}} B \log_{2} \left( 1 + \frac{ \left| p_{n} \right| ^{2} \left| \bar{h}_{n} \right| ^{2} }{ \sum_{i=1}^{n-1} \left| p_{i} \right| ^{2} \left| \bar{h}_{i} \right| ^{2} + \sigma^{2}} \right) \\
	{\rm s.t.}
	\ & {\rm (\ref{original_constraint_QoS}) \ and \ } \left| p_{n} \right| ^{2} \le P_{n}, \ \forall n \in \mathcal{N}.
	\end{align}
\end{subequations}

Note that the problem (\ref{transmit_problem_NOMA}) is not convex due to the non-convex objective function (\ref{transmit_problem_NOMA_objective}) and constraint (\ref{original_constraint_QoS}).
To handle the non-convexity of the objective function (\ref{transmit_problem_NOMA_objective}), we rewrite it as
\begin{align} \label{transmit_problem_NOMA_objective_rewrite}
R_{\rm NOMA}
& = \sum \nolimits_{n=K+1}^{K+N} B \log_{2} \left( \frac{ \sum_{i=1}^{n} \left| p_{i} \right| ^{2} \left| \bar{h}_{i} \right| ^{2} }{ \sum_{i=1}^{n-1} \left| p_{i} \right| ^{2} \left| \bar{h}_{i} \right| ^{2} + \sigma^{2}} \right) \nonumber \\
& = B \log_{2} \left( 1 + \frac{ \sum_{n=K+1}^{K+N} \left| p_{n} \right| ^{2} \left| \bar{h}_{n} \right| ^{2} }{ \sum_{k=1}^{K} \left| p_{k} \right| ^{2} \left| \bar{h}_{k} \right| ^{2} + \sigma^{2}} \right).
\end{align}

Next, we denote $\zeta_{n} = 2^{ R_{n}^{\min} / B} - 1, \forall n \in \mathcal{N}$, then the non-convex constraint (\ref{original_constraint_QoS}) can be transformed as
\begin{equation} \label{transmit_problem_NOMA_constraint_rewrite}
\left| p_{n} \right| ^{2} \left| \bar{h}_{n} \right| ^{2} \ge
\zeta_{n} \left( \sum \nolimits_{i = 1}^{n-1} \left| p_{i} \right| ^{2} \left| \bar{h}_{i} \right| ^{2} + \sigma^{2} \right), \ \forall n \in \mathcal{N}.
\end{equation}

Replacing the objective function (\ref{transmit_problem_NOMA_objective}) and constraint (\ref{original_constraint_QoS}) with the newly derived expressions in (\ref{transmit_problem_NOMA_objective_rewrite}) and (\ref{transmit_problem_NOMA_constraint_rewrite}), the non-convex problem (\ref{transmit_problem_NOMA}) can be rewritten as
\begin{subequations} \label{transmit_problem_NOMA_rewrite}
	\begin{align}
	\max \limits_{ \boldsymbol{p}_{\rm N}}
	\ & B \log_{2} \left( 1 + \frac{ \sum_{n=K+1}^{K+N} \left| p_{n} \right| ^{2} \left| \bar{h}_{n} \right| ^{2} }{ \sum_{k=1}^{K} \left| p_{k} \right| ^{2} \left| \bar{h}_{k} \right| ^{2} + \sigma^{2}} \right) \\
	{\rm s.t.}
	\ & {\rm (\ref{transmit_problem_NOMA_constraint_rewrite}) \ and \ } \left| p_{n} \right| ^{2} \le P_{n}, \ \forall n \in \mathcal{N}.
	\end{align}
\end{subequations}

Note that problem (\ref{transmit_problem_NOMA_rewrite}) is convex with $\rho_{n} = \left| p_{n} \right| ^{2}, \forall n \in \mathcal{N}$, thus it can be solved by the existing convex software packages such as CVX \cite{Grant2014CVX}.
After obtaining the solution $\rho_{n}^{*}$, the optimal transmit power at the $n$-th NOMA user can be recovered as $p_{n}^{*} = \sqrt{ \rho_{n}^{*} }, \forall n \in \mathcal{N}$.
Moreover, by taking a deep think for the power allocation problem (\ref{transmit_problem_NOMA_rewrite}), the closed-form solution to a simpler case is given in the following proposition.

\begin{proposition} \label{corollary_transmit_power}
	\emph{If all NOMA users do not have requirements on target data rate, i.e., $R_{n}^{\min} = 0, \forall n \in \mathcal{N}$, then each NOMA user just needs to simply perform full power transmission for the sum rate maximization \cite{Zeng2021Sum}, which can be given by}
	\begin{equation} \label{optimal_transmit_power_NOMA}
	p_{n}^{*} = \sqrt{ P_{n} }, \ \forall n \in \mathcal{N}.
	\end{equation}
\end{proposition}

\begin{IEEEproof}
	As one can see that the objective function is monotonically increasing with $p_{n}, \forall n \in \mathcal{N}$, thus the optimal power allocation at each NOMA user is to transmit at full power if constraint (\ref{transmit_problem_NOMA_constraint_rewrite}) is ignored.
	Moreover, it is worth pointing out that full power transmission is always the optimal solution for NOMA  users in the absence of QoS requirement, regardless of the configurations of the RIS in problem (\ref{transmit_problem_NOMA_rewrite}).
	This completes the proof.
\end{IEEEproof}

\subsubsection{Power allocation of AirFL users}
With obtained $\boldsymbol{p}_{\rm N}^{*}$, the power allocation problem of AirFL users is given by
\begin{subequations} \label{transmit_problem_AirFL}
	\begin{align}
	\label{transmit_problem_AirFL_objective}
	\max \limits_{ \boldsymbol{p}_{\rm A}}
	\ & (1 - \lambda) R_{\rm NOMA} + \lambda R_{\rm AirFL}  \\
	{\rm s.t.}
	\ & {\rm (\ref{original_constraint_MSE}) \ and \ } \left| p_{k} \right| ^{2} \le P_{k}, \ \forall k \in \mathcal{K}.
	\end{align}
\end{subequations}

In problem (\ref{transmit_problem_AirFL}), the MSE constraint (\ref{original_constraint_MSE}) and the power constraint are convex, but the objective function (\ref{transmit_problem_AirFL_objective}) is non-convex.
Thus, we define two convex functions as follows:
\begin{align}
\label{DC_convex_function_F}
F \left( \boldsymbol{p}_{\rm A} \right) 
& = (1 - \lambda) B \log_{2} \left( \sum \limits_{k=1}^{K} \left| p_{k} \right| ^{2} \left| \bar{h}_{k} \right| ^{2} + I_{\rm N} + \sigma^{2} \right) \nonumber \\
& + \lambda B \log_{2} \left( \sum \limits_{k=1}^{K} \left| a \bar{h}_k p_k \right| ^{2} + \left| a \right| ^{2} \sigma^{2} \right), \\
\label{DC_convex_function_G}
G \left( \boldsymbol{p}_{\rm A}, \beta \right) 
& = (1 - \lambda) B \log_{2} \left( \sum \nolimits_{k=1}^{K} \left| p_{k} \right| ^{2} \left| \bar{h}_{k} \right| ^{2} + \sigma^{2} \right) \nonumber \\
& + \lambda B \log_{2} \left( \beta + \left| a \right| ^{2} \sigma^{2} \right),
\end{align}
where $I_{\rm N} = \sum \nolimits_{n \in \mathcal{N}} \left| p_{n} \right| ^{2} \left| \bar{h}_{n} \right| ^{2}$ is a constant and $\beta = \sum \nolimits_{k \in \mathcal{K}} \left| a \bar{h}_{k} p_{k} - 1 \right| ^{2}$ is an auxiliary variable.

Then, the objective function (\ref{transmit_problem_AirFL_objective}) can be rewritten as the difference of two convex functions, i.e., $R_{\rm Hybrid} = F \left( \boldsymbol{p}_{\rm A} \right) - G \left( \boldsymbol{p}_{\rm A}, \beta \right)$.
As such, the problem (\ref{transmit_problem_AirFL}) can be re-expressed as a canonical DC programming problem with convex feasible set.
Specifically, by replacing (\ref{DC_convex_function_G}) with its first-order approximation, the non-convex objective function in (\ref{transmit_problem_AirFL_objective}) can be well approximated by the concave objective in (\ref{transmit_problem_AirFL_DC_objective}).
By doing so, the problem (\ref{transmit_problem_AirFL}) is reconstructed as a jointly convex optimization problem, which is given by
\begin{subequations} \label{transmit_problem_AirFL_DC}
	\begin{align}
	\label{transmit_problem_AirFL_DC_objective}
	\max \limits_{ \boldsymbol{p}_{\rm A}, \beta}
	\ & F \left( \boldsymbol{p}_{\rm A} \right) - \left\langle \nabla G ( \boldsymbol{p}_{\rm A}^{(\ell)} ),  \boldsymbol{p}_{\rm A}  \right\rangle - \left\langle \nabla G ( \beta^{(\ell)} ),  \beta \right\rangle \\
	{\rm s.t.}
	\label{beta_constraint}
	\ & \beta \ge \sum \nolimits_{k=1}^{K} \left| a \bar{h}_{k} p_{k} - 1 \right| ^{2}, \\
	\label{MSE_constraint}
	\ & \sum \nolimits_{k=1}^{K} \left| a \bar{h}_{k} p_{k} - 1 \right| ^{2} + \left| a \right| ^{2} \sigma^{2} \le \varepsilon_0 K^{2}, \\
	\ & \left| p_{k} \right| ^{2} \le P_{k}, \ \forall k \in \mathcal{K},
	\end{align}
\end{subequations}
where
$\boldsymbol{p}_{\rm A}^{(\ell)}$ and $\beta^{(\ell)}$ are the converged solutions after the $\ell$-th iteration,
$\nabla G ( \boldsymbol{p}_{\rm A}^{(\ell)} )$ and $\nabla G ( \beta^{(\ell)} )$ are the gradients of $G$ at points $\boldsymbol{p}_{\rm A}^{(\ell)}$ and $\beta^{(\ell)}$, respectively.
Therein, $\langle \boldsymbol{x}, \boldsymbol{y} \rangle$ denotes the dot product of two vector $\boldsymbol{x}$ and $\boldsymbol{y}$.
In fact, the DC programming is able to generate a sequence of improved feasible solutions with global convergence, which has been proved in \cite{Yang2020TWC}.
In summary, the DC-based algorithm for transmit power allocation is described in \textbf{Algorithm \ref{algorithm_DC_power_allocation}}.

\begin{algorithm}[t]
	\caption{DC-Based Algorithm for Power Allocation}
	\label{algorithm_DC_power_allocation}
	\begin{algorithmic}[1]
		\renewcommand{\algorithmicrequire}{\textbf{Initialize}}
		\renewcommand{\algorithmicensure}{\textbf{Output}}
		\STATE \textbf{Initialize} $\boldsymbol{p}^{(0)} = \left[ \boldsymbol{p}_{\rm A}^{(0)}, \boldsymbol{p}_{\rm N}^{(0)} \right] ^{\rm T}$, $\beta^{(0)}$, the tolerance $\epsilon_{1}$, the maximum iteration number $L_1$, and set $\ell_1=0$.
		\IF{$R_{n}^{\min} = 0, \forall n \in \mathcal{N}$}
		\STATE Obtain the optimal transmit power for NOMA users via (\ref{optimal_transmit_power_NOMA}), i.e., $\boldsymbol{p}_{\rm N}^{*} = \left[ \sqrt{P_{K+1}}, \sqrt{P_{K+2}}, \ldots, \sqrt{P_{K+N}}\right] ^{\rm T}$;
		\STATE Compute $U^{(\ell_1)} = R_{\rm Hybrid} \left( \boldsymbol{p}_{\rm A}^{(\ell_1)}, \boldsymbol{p}_{\rm N}^{*} \right)$;
			\REPEAT \label{DC_start}
			\STATE With given $\left( \boldsymbol{p}_{\rm A}^{(\ell_1)}, \beta^{(\ell_1)} \right)$, obtain $\left( \boldsymbol{p}_{\rm A}^{(\ell_1 + 1)}, \beta^{(\ell_1 + 1)} \right)$ by solving problem (\ref{transmit_problem_AirFL_DC}) to find the transmit power solution for AirFL users;
			\STATE Compute $U^{(\ell_1 + 1)} = R_{\rm Hybrid} \left( \boldsymbol{p}_{\rm A}^{(\ell_1 + 1)}, \boldsymbol{p}_{\rm N}^{*} \right)$;
			\STATE Update $\ell_1 := \ell_1 + 1$;
			\UNTIL $\left| U^{(\ell_1)} - U^{(\ell_1 - 1)} \right| \le \epsilon_1$ or $\ell_1 > L_1$; \label{DC_end}
		\ELSE
			\REPEAT
			\STATE With given $\boldsymbol{p}_{\rm A}^{(\ell_1)}$, obtain $\boldsymbol{p}_{\rm N}^{(\ell_1 + 1)}$ by solving problem (\ref{transmit_problem_NOMA_rewrite}) with CVX;
			\STATE With given $\boldsymbol{p}_{\rm N}^{(\ell_1 + 1)}$, repeat similar steps from \ref{DC_start} to \ref{DC_end} to obtain $\boldsymbol{p}_{\rm A}^{(\ell_1 + 1)}$;
			\STATE Compute $U^{(\ell_1 + 1)} = R_{\rm Hybrid} \left( \boldsymbol{p}_{\rm A}^{(\ell_1 + 1)}, \boldsymbol{p}_{\rm N}^{(\ell_1 + 1)} \right)$;
			\STATE Update $\ell_1 := \ell_1 + 1$;
			\UNTIL the solution improvement is below $\epsilon_1$ or the iteration number exceeds $L_1$;
		\ENDIF
		\STATE \textbf{Output} the transmit power solution $\boldsymbol{p}^{*} = \left[ \boldsymbol{p}_{\rm A}^{*}, \boldsymbol{p}_{\rm N}^{*} \right] ^{\rm T}$.
	\end{algorithmic}
\end{algorithm}

\subsection{Receive Scalar Control}
With the optimized $\boldsymbol{p}^{*}$ in problem (\ref{transceiver_problem}), the subproblem of receive scalar control is still intractable due to the non-convex objective function and constraint with respect to (w.r.t.) $a$, which can be equivalently expressed as
\begin{align} \label{receive_problem}
\max \limits_{ a } \ \log_{2} \left( \frac{ \sum \nolimits_{k=1}^{K} \left| a \bar{h}_k p_k \right| ^{2}  + \left| a \right| ^{2} \sigma^{2} }{ \sum \nolimits_{k=1}^{K} \left| a \bar{h}_k p_k - 1 \right| ^{2}  + \left| a \right| ^{2} \sigma^{2} } \right)
\quad {\rm s.t.} \ {\rm (\ref{original_constraint_MSE})}.
\end{align}

To simplify problem (\ref{receive_problem}), we introduce an auxiliary variable $\bar{a} = 1/ a$ with $a \ne 0$.
Then, problem (\ref{receive_problem}) is rewritten as
\begin{subequations} \label{receive_problem_rewritten}
	\begin{align}
	\label{receive_problem_objective_rewritten}
	\max \limits_{ \bar{a} }
	\ & \log_{2} \left( \frac{ \sum \nolimits_{k=1}^{K} \left| \bar{h}_k p_k \right| ^{2}  + \sigma^{2} }{ \sum \nolimits_{k=1}^{K} \left| \bar{h}_k p_k - \bar{a} \right| ^{2}  + \sigma^{2} } \right) \\
	{\rm s.t.}
	\label{receive_problem_MSE_constraint}
	\ & \sum \nolimits_{k=1}^{K} \left| \bar{h}_{k} p_{k} - \bar{a} \right| ^{2} + \sigma^{2} \le \varepsilon_0 K^{2} \left| \bar{a} \right| ^{2}.
	\end{align}
\end{subequations}

Note that the objective function (\ref{receive_problem_objective_rewritten}) is equivalent to minimize $\sum \nolimits_{k=1}^{K} \left| \bar{a} - \bar{h}_k p_k \right| ^{2}$, which is a convex function of $\bar{a}$.
To tackle the non-convexity of constraint (\ref{receive_problem_MSE_constraint}), the SCA method is invoked to replace $\left| \bar{a} \right| ^{2}$ with its first-order Taylor expansion.
The constraint (\ref{receive_problem_MSE_constraint}) is approximated by
\begin{align} \label{receive_problem_MSE_constraint_SCA}
& \sum \nolimits_{k=1}^{K} \left| \bar{h}_{k} p_{k} - \bar{a} \right| ^{2} + \sigma^{2} \nonumber \\
& \le \varepsilon_0 K^{2}
\left[ \left| \bar{a}^{(\ell)} \right| ^{2} + 2 {\rm Re} \left( \left( \bar{a}^{(\ell)} \right) ^{\rm H} \left( \bar{a} - \bar{a}^{(\ell)} \right)  \right) \right].
\end{align}

Based on the above approximations, the non-convex problem (\ref{receive_problem_MSE_constraint_SCA}) can be equivalently reformulated as the following convex one:
\begin{align} \label{receive_problem_SCA}
\min \limits_{ \bar{a} } \ \sum \nolimits_{k=1}^{K} \left| \bar{a} - \bar{h}_k p_k \right| ^{2}
\quad {\rm s.t.} \ {\rm (\ref{receive_problem_MSE_constraint_SCA})}.
\end{align}

Since problem (\ref{receive_problem_SCA}) is convex, the subproblem of receive scalar control can be efficiently addressed by using the SCA method, where the approximated convex problem (\ref{receive_problem_SCA}) is solved at each iteration to generate a sequence $\{ \bar{a}^{(\ell)} \}$ of improved feasible solutions.
With the converged $\bar{a}^{*}$, the optimal receive scalar can be recovered as $a^{*} = 1 / \bar{a}^{*}$.
The details of using SCA method to solve problem (\ref{receive_problem}) is presented in \textbf{Algorithm \ref{algorithm_SCA_receive_scalar}}.

To draw deep and useful insights into the receive scalar control problem in a simpler case, we consider that the AirFL users have no requirement on aggregation error, then the MSE constraint in (\ref{original_constraint_MSE}) can be ignored.
In that simplified case, there is no need to tackle the non-convex constraint (\ref{receive_problem_MSE_constraint}), and the receive scalar at the BS can be given in a closed-form expression, which is provided in the following proposition.

\begin{proposition} \label{corollary_receive_scalar}
	\emph{If all AirFL users do not have requirements on aggregation error, i.e., $\varepsilon_0 \rightarrow + \infty$, then the optimal receive scalar at the BS can be derived as
	\begin{equation} \label{optimal_receive_scalar}
	a^{*} = \frac{K}{ \sum \nolimits_{k=1}^{K} \bar{h}_k p_k },
	\end{equation}
	which is the reciprocal of the average value of the actual quality indicators (i.e., the product of the power budget and the combined channel gain) for all AirFL users.}
\end{proposition}

\begin{IEEEproof}
	When the MSE constraint in problem (\ref{receive_problem_rewritten}) is ignored, it reduces into an unconstrained optimization problem:
	\begin{equation} \label{receive_problem_unconstrained}
	\min \limits_{ \bar{a} } \ \sum \nolimits_{k=1}^{K} \left| \bar{a} - \bar{h}_k p_k \right| ^{2}.
	\end{equation}
	Suppose $f(\bar{a}) = \sum \nolimits_{k=1}^{K} \left( \bar{a} - \bar{h}_k p_k \right) ^{2}$, and set the first-order derivative of $f(\bar{a})$ to zero, i.e., $f'(\bar{a}) = 0$, the optimal solution of $\bar{a}$ can be obtained as
	\begin{equation}
	\bar{a}^{*} = \frac{ \sum \nolimits_{k=1}^{K} \bar{h}_k p_k }{K}.
	\end{equation}
	Thus, the optimal receive scalar can be recovered as $a^{*} = 1 / \bar{a}^{*}$, which completes the proof.
\end{IEEEproof}

\begin{algorithm}[t]
	\caption{SCA-Based Algorithm for Receive Scalar Control}
	\label{algorithm_SCA_receive_scalar}
	\begin{algorithmic}[1]
		\renewcommand{\algorithmicrequire}{\textbf{Initialize}}
		\renewcommand{\algorithmicensure}{\textbf{Output}}
		\STATE \textbf{Initialize} $a ^{(0)}$, the tolerance $\epsilon_2$, the maximum iteration number $L_2$, and set $\ell_2=0$.
		\IF{$\varepsilon_0 \rightarrow + \infty$}
		\STATE Obtain the optimal receive scalar $a^{*}$ at the BS via the closed-form solution in (\ref{optimal_receive_scalar}), i.e., $a^{*} = K / \sum \nolimits_{k=1}^{K} \bar{h}_k p_k$;
		\ELSE
			\STATE Compute $\bar{a} ^{(\ell_2)} = 1/ a ^{(\ell_2)}$ and $U^{(\ell_2)} = R_{\rm AirFL} \left( a ^{(\ell_2)} \right)$;
			\REPEAT
			\STATE Compute $\bar{a}^{(\ell_2 + 1)}$ by solving problem (\ref{receive_problem_SCA});
			\STATE Recover $a^{(\ell_2 + 1)} = 1 / \bar{a}^{(\ell_2 + 1)}$;
			\STATE Compute $U^{(\ell_2 + 1)} = R_{\rm AirFL} \left( a ^{(\ell_2 + 1)} \right)$;
			\STATE Update $\ell_2 := \ell_2 + 1$;
			\UNTIL $\left| U^{(\ell_2)} - U^{(\ell_2 - 1)} \right| \le \epsilon_2$ or $\ell_2 > L_2$;
		\ENDIF
		\STATE \textbf{Output} the receive scalar solution $a^{*}$. 
	\end{algorithmic}
\end{algorithm}

\subsection{Phase Shifts Design} \label{phase_shifts_design}
With obtained transceiver solutions $\boldsymbol{p}^{*}$ and $a^{*}$, the subproblem (\ref{RIS_problem}) w.r.t. discrete phase shifts is still challenging due to the combinatorial optimization nature.
As a result, we propose to first adopt the linear relaxation method to relax them into the continuous ones, i.e., $\theta_{m} \in \mathcal{A} \longrightarrow \theta_{m} \in [0, 2 \pi), \forall m \in \mathcal{M}$, and solve the relaxed problem with continuous variable $\boldsymbol{v}$\footnote{In order to simplify the expression and reduce the number of symbols, there are some symbol abuses in this subsection. For example, both the discrete phase shift and its continuous counterpart are denoted by $\theta_{m}$, so is $\boldsymbol{v}$. For clarity, we will put them in clear words when they represent discrete values.}, then the obtained phase-shifting solution $ \boldsymbol{\theta}^{*} = \left[ \theta_{1}, \theta_{2}, \ldots, \theta_{M} \right] $ would be recovered into their closest discrete values in $\mathcal{A}$.
Specifically, combining the definitions in (\ref{DC_convex_function_F}) and (\ref{DC_convex_function_G}), the continuous counterpart of subproblem (\ref{RIS_problem}) can be written as
\begin{subequations} \label{RIS_problem_continuous}
	\begin{align}
	\label{RIS_problem_continuous_objective}
	\max \limits_{ \boldsymbol{v} }
	\ & F \left( \boldsymbol{v} \right)  - G \left( \boldsymbol{v} \right) \\
	{\rm s.t.}
	\label{RIS_problem_continuous_constraints}
	\ & {\rm (\ref{state_factor_order}), (\ref{original_constraint_QoS}) \ and \ (\ref{original_constraint_MSE})}, \\
	\ & \theta_{m} \in [0, 2 \pi), \ \forall m \in \mathcal{M},
	\end{align}
\end{subequations}
where all constraints in (\ref{RIS_problem_continuous_constraints}) represent their continuous counterparts.
Nevertheless, the problem is still non-convex due to the non-convex objective function and constraints.

To facilitate derivation, we set $\alpha=2$, the combined channel coefficient is then given by\footnote{We assume that the small-scale fading $\boldsymbol{g}$ and $\boldsymbol{h}_{i}$ are only related to the statistical channel information, but not related to the location of RIS \cite{Mu2020Joint}, thus $\boldsymbol{\Phi}_{i} = {\rm diag} ( \boldsymbol{g}^{\rm H} ) \boldsymbol{h}_{i}$ is assumed to be unchanged during one transmission.}
\begin{equation} \label{hat_h}
\bar{h}_{i} = \frac{\varsigma_0 \boldsymbol{v}^{\rm H} \boldsymbol{\Phi}_{i}}{d_{0} d_{i}}
= \boldsymbol{v}^{\rm H} \widetilde{ \boldsymbol{\Phi} }_{i}, \ \forall i \in \mathcal{I},
\end{equation}
where $\widetilde{ \boldsymbol{\Phi} }_{i} = \frac{\varsigma_0 \boldsymbol{\Phi}_{i}}{d_{0} d_{i}}, \forall i \in \mathcal{I}$.
Besides, we give the following three lemma to tackle the non-convexity of problem (\ref{RIS_problem_continuous}).

\begin{lemma} \label{lemma_1}
	\emph{If we define $\boldsymbol{\Lambda}_{i} = \widetilde{ \boldsymbol{\Phi} }_{i} \widetilde{ \boldsymbol{\Phi} }_{i}^{ \rm H }$, $ \widetilde{ \boldsymbol{V} } = \boldsymbol{v} \boldsymbol{v}^{\rm H}$ and $\widetilde{ \boldsymbol{V} } \succeq \boldsymbol{0}$, then the combined channel gain is a convex function w.r.t. $\widetilde{ \boldsymbol{V} }$, i.e.,}
	\begin{equation} \label{lemma_trace_V}
	\left| \bar{h}_{i} \right| ^2 = {\rm tr} \left( \boldsymbol{\Lambda}_{i} \widetilde{ \boldsymbol{V} } \right), \ \forall i \in \mathcal{I}.
	\end{equation}
\end{lemma}

\begin{IEEEproof}
	Please refer to Appendix \ref{proof_of_lemma_1}.
\end{IEEEproof}

\begin{lemma} \label{lemma_2}
	\emph{If we define $\widehat{ \boldsymbol{\Phi}}_{k} = \frac{ a \varsigma_0 \boldsymbol{\Phi}_{k} p_{k} }{ d_{0} d_{k} }, \forall k \in \mathcal{K}$ and $\widetilde{ \boldsymbol{\Lambda} }_{k} = \widehat{ \boldsymbol{\Phi}}_{k} \widehat{ \boldsymbol{\Phi}}_{k}^{\rm H}, \forall k \in \mathcal{K}$, then we have}
	\begin{equation} \label{lemma_trace_tilde_V}
	\left| a \bar{h}_{k} p_{k} \right| ^{2}
	= {\rm tr} \left( \widetilde{ \boldsymbol{\Lambda} }_{k} \widetilde{ \boldsymbol{V} } \right), \ \forall k \in \mathcal{K}.
	\end{equation}
\end{lemma}

\begin{IEEEproof}
	Please refer to Appendix \ref{proof_of_lemma_2}.
\end{IEEEproof}

\begin{lemma} \label{lemma_3}
	\emph{If we define
		\begin{equation}
		\widehat{ \boldsymbol{\Lambda}}_{k} = \left[ \begin{array}{cc}
		\widehat{ \boldsymbol{\Phi}}_{k} \widehat{ \boldsymbol{\Phi}}_{k}^{\rm H} & - \widehat{ \boldsymbol{\Phi}}_{k} \\
		- \widehat{ \boldsymbol{\Phi}}_{k}^{\rm H} & 0 
		\end{array} 
		\right ]
		\text{and} \
		\boldsymbol{\bar{v}} = \left[ \begin{array}{c}
		\boldsymbol{v} \\
		1
		\end{array} 
		\right ],
		\end{equation}
		then we have
		\begin{equation} \label{lemma_trace_hat_V}
		\left| a \bar{h}_{k} p_{k} - 1 \right| ^{2}
		= {\rm tr} \left( \widehat{ \boldsymbol{\Lambda}}_{k} \boldsymbol{V} \right) + 1, \ \forall k \in \mathcal{K},
		\end{equation}
		where $\boldsymbol{V} = \boldsymbol{\bar{v}} \boldsymbol{\bar{v}}^{\rm H}$, while $\boldsymbol{V} \succeq \boldsymbol{0}$ and ${\rm rank} ( \boldsymbol{V} ) = 1$.}
\end{lemma}

\begin{IEEEproof}
	Please refer to Appendix \ref{proof_of_lemma_3}.
\end{IEEEproof}

To unify the variables in (\ref{lemma_trace_V}), (\ref{lemma_trace_tilde_V}) and (\ref{lemma_trace_hat_V}) as $\boldsymbol{V}$, the expressions in (\ref{lemma_trace_V}) and (\ref{lemma_trace_tilde_V}) can be rewritten as
\begin{align}
	\label{trace_V_rewrite_I}
	\left| \bar{h}_{i} \right| ^2
	& = {\rm tr} \left( \mathring{ \boldsymbol{\Lambda} }_{i} \boldsymbol{v} \boldsymbol{v}^{\rm H} \right)
	= {\rm tr} \left( \mathring{ \boldsymbol{\Lambda} }_{i} \boldsymbol{V} \right), \ \forall i \in \mathcal{I}, \\
	\label{trace_V_rewrite_K}
	\left| a \bar{h}_{k} p_{k} \right| ^{2}
	& = {\rm tr} \left( \ddot{ \boldsymbol{\Lambda} }_{k} \boldsymbol{v} \boldsymbol{v}^{\rm H} \right)
	= {\rm tr} \left( \ddot{ \boldsymbol{\Lambda} }_{k} \boldsymbol{V} \right), \ \forall k \in \mathcal{K}.
\end{align}
where $\mathring{ \boldsymbol{\Lambda} }_{i} = {\rm diag} ( \widetilde{ \boldsymbol{\Phi} }_{i} \widetilde{ \boldsymbol{\Phi} }_{i}^{ \rm H }, 0 ), \forall i $ and $\ddot{ \boldsymbol{\Lambda} }_{k} = {\rm diag} ( \widehat{ \boldsymbol{\Phi} }_{k} \widehat{ \boldsymbol{\Phi} }_{k}^{ \rm H }, 0 ), \forall k$ are obtained by matrix lifting.

\begin{figure*}[t]
	\begin{equation} \label{F_V}
	F \left( \boldsymbol{V} \right) = (1 \!-\! \lambda) \!B\! \log_{2} \left( \sum \limits_{ i \in \mathcal{I} } \left| p_{i} \right| ^{2} {\rm tr} \left( \mathring{ \boldsymbol{\Lambda} }_{i} \boldsymbol{V} \right) \!+\! \sigma^{2} \right) \!+\! \lambda \!B\! \log_{2} \left( \sum \limits_{ k \in \mathcal{K} } {\rm tr} \left( \ddot{ \boldsymbol{\Lambda} }_{k} \boldsymbol{V} \right) \!+\! \left| a \right| ^{2} \sigma^{2} \right)
	\end{equation}
	\begin{equation} \label{G_V}
	G \left( \boldsymbol{V} \right) = (1 \!-\! \lambda) \!B\! \log_{2} \left( \sum \limits_{ k \in \mathcal{K} } \left| p_{k} \right| ^{2} {\rm tr} \left( \mathring{ \boldsymbol{\Lambda} }_{k} \boldsymbol{V} \right) \!+\! \sigma^{2} \right) \!+\! \lambda \!B\! \log_{2} \left( \sum \limits_{ k \in \mathcal{K} } {\rm tr} \left( \widehat{ \boldsymbol{\Lambda}}_{k} \boldsymbol{V} \right) \!+\! K \!+\! \left| a \right| ^{2} \sigma^{2} \right)
	\end{equation}
	\hrulefill
\end{figure*}

Based on the above derivations in (\ref{lemma_trace_hat_V}), (\ref{trace_V_rewrite_I}) and (\ref{trace_V_rewrite_K}),
the hybrid rate can be rewritten as $R_{\rm Hybrid} = F \left( \boldsymbol{V} \right) - G \left( \boldsymbol{V} \right)$, where the exact expressions of $F \left( \boldsymbol{V} \right)$ and $G \left( \boldsymbol{V} \right)$ are given in (\ref{F_V}) and (\ref{G_V}).
In addition, the non-convex constraints (\ref{state_factor_order}), (\ref{original_constraint_QoS}) and (\ref{original_constraint_MSE}) can be transformed as the following convex ones:
{\allowdisplaybreaks[4]
\begin{align}
\label{V_state_factor_constraint}
{\rm tr} \left( \mathring{ \boldsymbol{\Lambda} }_{k} \boldsymbol{V} \right)
\le
{\rm tr} \left( \mathring{ \boldsymbol{\Lambda} }_{K+1} \boldsymbol{V} \right)
\le \ldots \le
{\rm tr} \left( \mathring{ \boldsymbol{\Lambda} }_{K+N} \boldsymbol{V} \right), \\
\label{V_QoS_constraint}
\left| p_{n} \right| ^{2} {\rm tr} \left( \mathring{ \boldsymbol{\Lambda} }_{n} \boldsymbol{V} \right)
\ge
\zeta_{n} \left( \sum \limits_{i = 1}^{n-1} \left| p_{i} \right| ^{2} {\rm tr} \left( \mathring{ \boldsymbol{\Lambda} }_{i} \boldsymbol{V} \right) + \sigma^{2} \right), \\
\label{V_MSE_constraint}
\sum \nolimits_{k=1}^{K} {\rm tr} \left( \widehat{ \boldsymbol{\Lambda}}_{k} \boldsymbol{V} \right) + K + \left| a \right| ^{2} \sigma^{2} \le \varepsilon_0 K^{2}.
\end{align}


With the newly derived objective function and constraints, the continuous phase shifts design problem (\ref{RIS_problem_continuous}) can be expressed as}
{\allowdisplaybreaks[4]
\begin{subequations} \label{RIS_problem_reflection}
	\begin{align}
	\label{RIS_problem_reflection_objective}
	\max \limits_{ \boldsymbol{V} }
	\ & F \left( \boldsymbol{V} \right)  - G \left( \boldsymbol{V} \right) \\
	{\rm s.t.}
	\ & {\rm (\ref{V_state_factor_constraint}), (\ref{V_QoS_constraint}) \ and \ (\ref{V_MSE_constraint})}, \\
	\label{V_m_1_constraint}
	\ & \boldsymbol{V}_{m,m} = 1, \ \forall m = 1,2,\ldots,M+1, \\
	\label{V_0_constraint}
	\ &	\boldsymbol{V} \succeq \boldsymbol{0}, \\
	\label{rank_one_constraint}
	\ & \bcancel{{\rm rank} ( \boldsymbol{V} ) = 1}.
	\end{align}
\end{subequations}

Since only the rank-one constraint (\ref{rank_one_constraint}) is non-convex in problem (\ref{RIS_problem_reflection}), the SDR method can be invoked to relax this constraint.
After that, the relaxed problem of (\ref{RIS_problem_reflection}) is a convex semidefinite programming (SDP) problem, thus the optimal solution $\boldsymbol{V}^{*}$ can be obtained by using the well-known toolbox such as CVX.}
Note that if the obtained solution $\boldsymbol{V}^{*}$ fails to be rank-one, the Gaussian randomization method \cite{Luo2010Semidefinite} can be adopted to obtain a reconstructed rank-one solution to problem (\ref{RIS_problem_reflection}).
Alternatively,  in order to avoid the limitations brought by ignoring the rank-one constraint directly, the sequential rank-one constraint relaxation approach \cite{Mu2020Exploiting} can also be invoked to solve problem (\ref{RIS_problem_reflection}) for finding a local optimal rank-one solution.

Eventually, by using the Cholesky decomposition $\boldsymbol{V}^{*} = \boldsymbol{\bar{v}}^{*} \boldsymbol{\bar{v}}^{* \rm H}$ and $\boldsymbol{\bar{v}}^{*} = \left[ \boldsymbol{v}^{*}, 1 \right] ^{\rm T}$, the optimal continuous reflection vector $\boldsymbol{v}^{*}$ can be obtained.
Afterwards, the discrete phase shifts can be straightforwardly recovered as
\begin{equation} \label{RIS_problem_discrete_phase_recover}
\theta_{m} = \arg \min_{\theta \in \mathcal{A}} \left| \theta - \theta_{m}^{*} \right|, \ \forall m \in \mathcal{M}. 
\end{equation}
where $\theta_{m}^{*} = {\rm arg} ( \boldsymbol{v}_{m}^{*} )$ is the continuous phase shift of the $m$-th element of the obtained reflection vector.
Based on the above results, the SDR-based algorithm for discrete phase shifts design can be summarized in \textbf{Algorithm \ref{algorithm_SDR_phase_shifts}}.

\begin{remark}
	\emph{The proposed relaxation-then-quantization scheme is a direct approach to find a feasible solution, and is highly suitable for the high-resolution case or the large RIS case that equips with many reflecting elements, especially for the extreme case with continuous phase shifts.
	However, it may be ineffective for some practical cases with very low resolution (e.g., $b=1$) and small RIS (e.g., $M \le 5$), due to the non-negligible performance loss.
	To overcome this drawback, an exhaustive search-based method can be used to obtain an optimal solution for the low resolution and small RIS cases.}
\end{remark}

\begin{algorithm}[t]
	\caption{SDR-Based Algorithm for Phase Shifts Design}
	\label{algorithm_SDR_phase_shifts}
	\begin{algorithmic}[1]
		\renewcommand{\algorithmicrequire}{\textbf{Initialize}}
		\renewcommand{\algorithmicensure}{\textbf{Output}}
		\STATE \textbf{Initialize} $\boldsymbol{v} ^{(0)}$, the tolerance $\epsilon_3$, the maximum iteration number $L_3$, and set $\ell_3=0$.
		\IF{both $b$ and $M$ are small}
		\STATE Solve problem (\ref{RIS_problem}) via an exhaustive search method;
		\ELSE
		\STATE Relax the phase shifts as $\theta_{m} \in [0, 2 \pi), \forall m \in \mathcal{M}$;
		\STATE Compute $\boldsymbol{\bar{v}}^{(\ell_3)} = [ \boldsymbol{v}^{(\ell_3)}, 1 ] ^{\rm T}$ and $\boldsymbol{V}^{(\ell_3)} = \boldsymbol{\bar{v}}^{(\ell_3)} \left( \boldsymbol{\bar{v}}^{(\ell_3)} \right) ^{\rm H}$;
		\STATE Compute $U^{(\ell_3)} = F \left( \boldsymbol{V}^{(\ell_3)} \right)  - G \left( \boldsymbol{V}^{(\ell_3)} \right)$;
		\REPEAT \label{SDP_start}
		\STATE With given $\boldsymbol{V}^{(\ell_3)}$, obtain $\boldsymbol{V} ^{(\ell_3 + 1)}$ by solving the relaxed SDP problem of (\ref{RIS_problem_reflection});
		\STATE Compute $U^{(\ell_3 + 1)} = F \left( \boldsymbol{V}^{(\ell_3 + 1)} \right)  - G \left( \boldsymbol{V}^{(\ell_3 + 1)} \right)$;
		\STATE Update $\ell_3 := \ell_3 + 1$;
		\UNTIL $\left| U^{(\ell_3)} - U^{(\ell_3 - 1)} \right| \le \epsilon_3$ or $\ell_3 > L_3$; \label{SDP_end}
		\STATE Obtain $\boldsymbol{\bar{v}}^{*}$ by Cholesky decomposition $\boldsymbol{V}^{*} = \boldsymbol{\bar{v}}^{*} \boldsymbol{\bar{v}}^{* \rm H}$;
		\STATE Obtain $\boldsymbol{v}^{*}$ according to $\boldsymbol{\bar{v}}^{*} = \left[ \boldsymbol{v}^{*}, 1 \right] ^{\rm T}$;
		\STATE Recover the discrete phase shifts $\boldsymbol{\theta}^{*}$ via (\ref{RIS_problem_discrete_phase_recover});
		\ENDIF
		\STATE \textbf{Output} the phase-shifting solution $\boldsymbol{\theta}^{*} = \left[ \theta_{1}, \theta_{2}, \ldots, \theta_{M} \right] $.
	\end{algorithmic}
\end{algorithm}

\subsection{Convergence and Complexity Analysis} \label{convergence_complexity}
Invoking the designed algorithms to find suboptimal solutions for the corresponding decomposed subproblems, an alternating optimization algorithm for solving the MINLP problem (\ref{original_problem}) is given in \textbf{Algorithm \ref{algorithm_alternating_optimization}}.
In the first step, the transmit power at each user is performed based on the DC programming-based allocation strategy, i.e., \textbf{Algorithm \ref{algorithm_DC_power_allocation}}.
In the second step, the receive scalar at the BS is controlled by the converged solution obtained from the SCA-based receive scaling approach, i.e., \textbf{Algorithm \ref{algorithm_SCA_receive_scalar}}.
In the third step, the design of the discrete phase shifts is determined according to the relaxation-then-quantization method, i.e., \textbf{Algorithm \ref{algorithm_SDR_phase_shifts}}.

In addition, when the NOMA users have no requirement on the target data rate, i.e., $R_{n}^{\min} = 0, \forall n \in \mathcal{N}$, they just need to perform full power transmission described in \textit{Proposition \ref{corollary_transmit_power}}.
On the other hand, if the AirFL users can tolerate a large error, i.e., $\varepsilon_0 \gg 0$, the optimal receive scalar can be controlled by the closed-form solution in \textit{Proposition \ref{corollary_receive_scalar}}.
In these simpler cases, the complexity of \textbf{Algorithm \ref{algorithm_alternating_optimization}} can be largely reduced.
Moreover, the convergence and complexity of the three-step alternating optimization algorithm are analyzed in the following context.

\begin{algorithm}[t]
	\caption{AO-Based Algorithm for Solving Problem (\ref{original_problem})}
	\label{algorithm_alternating_optimization}
	\begin{algorithmic}[1]
		\renewcommand{\algorithmicrequire}{\textbf{Initialize}}
		\renewcommand{\algorithmicensure}{\textbf{Output}}
		\STATE \textbf{Initialize} $( \boldsymbol{p}^{(0)}, a^{(0)}, \boldsymbol{v} ^{(0)} )$, the tolerance $\epsilon_4$, the maximum iteration number $L_4$, and set $\ell_4=0$.
		\STATE Compute $U^{(\ell_4 + 1)} = R_{\rm Hybrid} \left( \boldsymbol{p}^{(\ell_4)}, a^{(\ell_4)}, \boldsymbol{v} ^{(\ell_4)} \right)$;
		\REPEAT
		\STATE \textbf{Step 1:} Transmit power allocation
		\STATE Given $( a^{(\ell_4)}, \boldsymbol{v} ^{(\ell_4)} )$, obtain $\boldsymbol{p}^{(\ell_4+1)}$ by solving problems (\ref{transmit_problem_NOMA}) and (\ref{transmit_problem_AirFL}) via Algorithm \ref{algorithm_DC_power_allocation}.
		\STATE \textbf{Step 2:} Receive scalar control
		\STATE Given $( \boldsymbol{p}^{(\ell_4+1)}, \boldsymbol{v} ^{(\ell_4)} )$, obtain $a^{(\ell_4+1)}$ by solving problem (\ref{receive_problem}) via Algorithm \ref{algorithm_SCA_receive_scalar}.
		\STATE \textbf{Step 3:} Phase shifts design
		\STATE Given $( \boldsymbol{p}^{(\ell_4+1)}, a^{(\ell_4+1)} )$, obtain $( \boldsymbol{v} ^{(\ell_4 + 1)} )$ by solving problem (\ref{RIS_problem}) via Algorithm \ref{algorithm_SDR_phase_shifts};
		\STATE Update $\ell_4 := \ell_4 + 1$;
		\UNTIL $\left| U^{(\ell_4)} - U^{(\ell_4 - 1)} \right| \le \epsilon_4$ or $n_{4} > N_{4}$;
		\STATE \textbf{Output} the converged solution $( \boldsymbol{p}^{*}, a^{*}, \boldsymbol{v} ^{*} )$. 
	\end{algorithmic}
\end{algorithm}

\subsubsection{Convergence}
In \textbf{Algorithm \ref{algorithm_alternating_optimization}}, we denote $(  \boldsymbol{p}^{(\ell)}, a^{(\ell)}, \boldsymbol{v} ^{(\ell)} )$ as the solution to problem (\ref{original_problem}) obtained in the $\ell$-th iteration, where the objective value is defined as
\begin{equation} \label{objective_value}
U \left(  \boldsymbol{p}^{(\ell)}, a^{(\ell)}, \boldsymbol{v} ^{(\ell)} \right) = R_{\rm Hybrid} \left(  \boldsymbol{p}^{(\ell)}, a^{(\ell)}, \boldsymbol{v} ^{(\ell)} \right) .
\end{equation}

Let $U ^{(\ell)} = U \left( \boldsymbol{p}^{(\ell)}, a^{(\ell)}, \boldsymbol{v} ^{(\ell)} \right)$, and substitute the current solution $(  \boldsymbol{p}^{(\ell)}, a^{(\ell)}, \boldsymbol{v} ^{(\ell)} )$ into (\ref{original_problem}).
Then, executing \textbf{Step 1-2-3} once again, we have
\begin{align} \label{iteration_value}
U ^{(\ell)} & = U \left(  \boldsymbol{p}^{(\ell)}, a^{(\ell)}, \boldsymbol{v} ^{(\ell)} \right)
\overset{(a)}{\le} U \left( \boldsymbol{p}^{(\ell +1)}, a^{(\ell)}, \boldsymbol{v} ^{(\ell)} \right) \nonumber \\
& \overset{(b)}{\le} U \left( \boldsymbol{p}^{(\ell +1)}, a^{(\ell +1)}, \boldsymbol{v} ^{(\ell)} \right)
\! \overset{(d)}{\le} \! U \left( \boldsymbol{p}^{(\ell +1)}, a^{(\ell +1)}, \boldsymbol{v} ^{(\ell +1)} \right) \nonumber \\
& = U ^{(\ell+1)},
\end{align}
where the inequality (a), (b), and (c) comes from the fact that the continuous improvements of the transmit power allocation in \textbf{Step 1}, the receive scalar control in \textbf{Step 2}, and the combinatorial optimization of reflection design in \textbf{Step 3}, respectively.
Therefore, the achievable hybrid rate is non-decreasing over the iterations.
Meanwhile, owing to the finite power budget and bandwidth, the optimal hybrid rate has a limited upper bound.
Thus, the sequence $\{ U ^{(\ell)} \}$ can at least converge to a locally optimal solution of the original MINLP problem (\ref{original_problem}), if not an optimal one.

\subsubsection{Complexity}
When the convex subproblems are solved by CVX, the interior point method is considered, unless otherwise stated.
For \textbf{Algorithm \ref{algorithm_alternating_optimization}}, the main complexity of solving problem (\ref{original_problem}) lies in addressing the power allocation problems with \textbf{Algorithm \ref{algorithm_DC_power_allocation}} and dealing with the reflection design problem with \textbf{Algorithm \ref{algorithm_SDR_phase_shifts}}.
Specifically, the worst-case complexity of \textbf{Algorithm \ref{algorithm_DC_power_allocation}} can be expressed as $\mathcal{O}\left( L_1 N^3 + L_1^2 (K+1)^3 \right)$.
The complexity of \textbf{Algorithm \ref{algorithm_SDR_phase_shifts}} mainly depends on the steps from \ref{SDP_start} to \ref{SDP_end} for solving the convex SDP problem (\ref{RIS_problem_reflection}), which can be given by $\mathcal{O}\left( L_3 M^6 \right)$.
Thus,  the overall complexity of solving the MINLP problem (\ref{original_problem}) with \textbf{Algorithm \ref{algorithm_SDR_phase_shifts}} can be represented as $\mathcal{O} \left(  L_1 L_4 N^3 + L_1^2 L_4 (K+1)^3 + L_3 L_4 M^6 \right) $, which can be further reduced for these simplified cases without QoS and/or MES constraints.

\subsection{Extension to Multi-Antenna BS and Imperfect SIC Scenario}
When the BS has $N_r$ antennas, the BS-RIS link is denoted by $\boldsymbol{G} \in \mathbb{C}^{M \times N_r}$, and the receive vector at the BS is denoted by $\boldsymbol{a} \in \mathbb{C}^{1 \times N_r}$.
Let $\boldsymbol{\tilde{\Phi}}_{i} = \boldsymbol{G}^{\rm H} {\rm diag} ( \boldsymbol{h}_{i} ) $. Then, the combined channel coefficient is given by
$\boldsymbol{\bar{h}}_{i} =  \varsigma_0 \sqrt{(d_{0} d_{i})^{-\alpha}} \boldsymbol{\tilde{\Phi}}_{i} \boldsymbol{v}, \forall i \in \mathcal{I}$.
The superposition signal received at the BS is expressed as
\begin{equation}
\setlength\abovedisplayskip{3pt}
\setlength\belowdisplayskip{3pt}
\boldsymbol{y} = \underbrace{\sum \nolimits_{k=1}^{K} \boldsymbol{\bar{h}}_{k} p_{k} s_{k}}_{\mathbf {AirFL~users}}
+ \underbrace{\sum \nolimits_{n=K+1}^{K+N} \boldsymbol{\bar{h}}_{n} p_{n} s_{n}}_{\mathbf {NOMA~users}}
+ \underbrace{\boldsymbol{z}_{0}}_{\mathbf {noise}},
\end{equation}
where $\boldsymbol{z}_{0} \sim \mathcal{CN} (0, \sigma^{2} \boldsymbol{I})$ is the AWGN at the BS.

Similar to the decoding order (\ref{state_factor_order}) in the single-antenna BS scenario, the SIC decoding order in the multi-antenna BS scenario can be ranked as
\begin{equation} \label{state_factor_order_E3}
\setlength\abovedisplayskip{3pt}
\setlength\belowdisplayskip{3pt}
\underbrace{ \left\| \boldsymbol{\bar{h}}_{k} \right\| ^{2} \le \left\| \boldsymbol{\bar{h}}_{K+1} \right\| ^{2} \cdots \le }_{\mathbf {weak~signals~that~can~not \atop be~cancelled}}
\left\| \boldsymbol{\bar{h}}_{n} \right\| ^{2}
\underbrace{ \cdots \le \left\| \boldsymbol{\bar{h}}_{K+N} \right\| ^{2},}_{\mathbf {strong~signals~that~can \atop be~partially~cancelled}}
\ \forall k, n. 
\end{equation}

Accordingly, a generic SINR representation of the imperfect SIC receiver can be given by
\begin{equation} \label{gamma_E4}
\bar{\gamma}_{n} = \frac{ \left| p_{n} \right| ^{2} \left\| \boldsymbol{\bar{h}}_{n} \right\| ^{2} }{ \sum\limits_{k=1}^{n-1} \left| p_{k} \right| ^{2} \left\| \boldsymbol{\bar{h}}_{k} \right\| ^{2} + \epsilon_b \sum\limits_{i=n+1}^{N+K} \left| p_{i} \right| ^{2} \left\| \boldsymbol{\bar{h}}_{i} \right\| ^{2} + \sigma^{2} }, \ \forall n \in \mathcal{N},
\end{equation}
where $\epsilon_b \in [0, 1]$ denotes the imperfect SIC coefficient that characterizes the residual interference from strong users.
Specifically, $\epsilon_b=0$ is the ideal case of perfect SIC, while $\epsilon_b=1$ represents the situation without SIC.
Based on (\ref{gamma_E4}), the sum rate of all NOMA users is given by
\begin{equation} \label{NOMA_rate_E5}
\bar{R}_{\rm NOMA} = \sum \nolimits_{n=K+1}^{K+N} 
B \log_{2} \left( 1 + \bar{\gamma}_{n} \right).
\end{equation}

Due to the imperfect SIC, the residual signal for AirFL model aggregation is given by
$\boldsymbol{\hat{y}} = \sum \nolimits_{k=1}^{K} \boldsymbol{\bar{h}}_{k} p_{k} s_{k} + \sqrt{\epsilon_b} \sum \nolimits_{n=K+1}^{K+N} \boldsymbol{\bar{h}}_{n} p_{n} s_{n} + \boldsymbol{z}_{0}$.
By applying the receive vector $\boldsymbol{a}$ to the residual signal $\boldsymbol{\hat{y}}$, we get the reconstructed AirFL signal as
$\bar{s} = \boldsymbol{a} \boldsymbol{\hat{y}} / K$.
Then, the MSE in the multi-antenna BS and imperfect SIC scenario is given by
\begin{align} \label{MSE_s_E8}
\setlength\abovedisplayskip{3pt}
\setlength\belowdisplayskip{3pt}
& {\rm MSE} (\bar{s}, s) \triangleq \mathbb{E} ( |\bar{s} - s|^2) \nonumber \\
& = \frac{1}{K^{2}} \sum_{k=1}^{K} \left| \boldsymbol{a} \boldsymbol{\bar{h}}_{k} p_{k} - 1 \right| ^2
+ \frac{\epsilon_b}{K^{2}} \sum_{n=K+1}^{K+N} \left| \boldsymbol{a} \boldsymbol{\bar{h}}_{n} p_{n} \right| ^2
+ \frac{ \| \boldsymbol{a} \|^{2} \sigma^2}{K^{2}}.
\end{align}

\begin{figure*}
	\begin{equation} \label{AirFL_rate_E9}
	\bar{R}_{\rm AirFL} 
	= B \log_{2} \left( \frac{ \sum_{k=1}^{K} \left| \boldsymbol{a} \boldsymbol{\bar{h}}_{k} p_{k} \right| ^2
		+ \epsilon_b \sum \nolimits_{n=K+1}^{K+N} \left| \boldsymbol{a} \boldsymbol{\bar{h}}_{n} p_{n} \right| ^2
		+  \| \boldsymbol{a} \|^{2} \sigma^2 }{  \sum_{k=1}^{K} \left| \boldsymbol{a} \boldsymbol{\bar{h}}_{k} p_{k} - 1 \right| ^2
		+ \epsilon_b \sum \nolimits_{n=K+1}^{K+N} \left| \boldsymbol{a} \boldsymbol{\bar{h}}_{n} p_{n} \right| ^2
		+ { \| \boldsymbol{a} \|^{2} \sigma^2} } \right)
	\end{equation}
	\hrule
\end{figure*}

Similar to (\ref{AirFL_rate}) in Definition \ref{definition_AirFL_rate}, the computation rate of AirFL users is given in (\ref{AirFL_rate_E9}) at the top of the next page.
Finally, substituting (\ref{state_factor_order_E3})-(\ref{AirFL_rate_E9}) into problem (\ref{original_problem}), we can obtain the optimization problem for the multi-antenna BS and imperfect SIC scenario.
Due to the problem similarity, the decomposition method of Section \ref{section_problem_formulation} and the iterative algorithms of Section \ref{alternating_optimization} can be extended to solve it efficiently, details of which are omitted for simplicity.

\section{Numerical Results} \label{section_simulation}

\begin{figure} [t!]
	\centering
	\includegraphics[width=3.5 in]{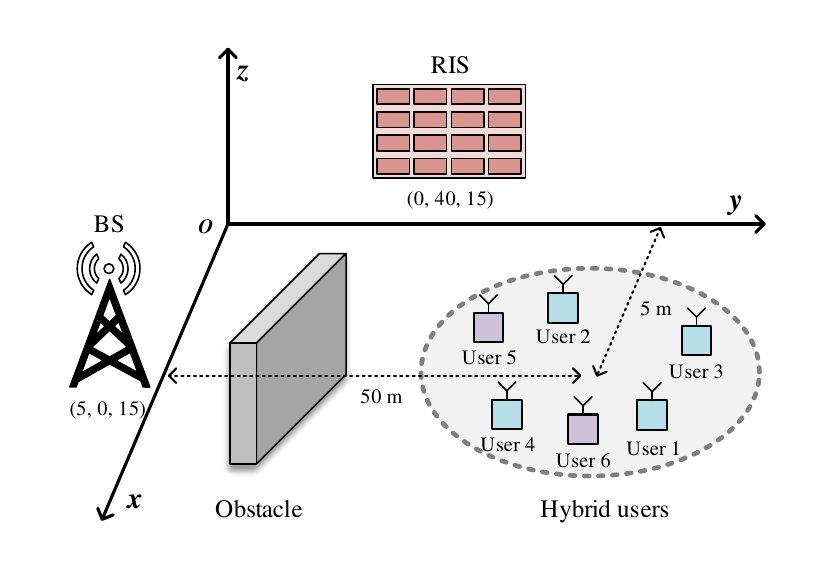}
	\caption{Simulation setup of the RIS-aided hybrid network (3D view).}
	\label{simulation_setup}
\end{figure}

\subsection{Simulation Settings}
The simulation setup is shown in Fig. \ref{simulation_setup}, we consider that there are $I=6$ users and one BS in the RIS-aided hybrid network, where $K=4$ AirFL users and $N=2$ NOMA users are randomly and uniformly distributed in a circle centered at $(5, 50, 0)$ (in meters) with radius ${\rm  3 \ m}$.
In the three-dimensional (3D) Cartesian coordinates, the BS and RIS are located at $(5, 0, 15)$ and $(0, 40, 15)$, respectively.
Moreover, the reference pass loss is set as $\varsigma_0 = -30 {\rm \ dBm}$, and the Rician factor is $2$.
The power budget of the $i$-th user is $P_{i}=P_{\max}=23 {\rm \ dBm}, \forall i \in \mathcal{I}$, and the noise power is $\sigma^{2} = -80 {\rm \ dBm}$.
The transmission bandwidth is set as $B = 1 {\rm \ MHz}$, the minimum rate requirement of the $n$-th NOMA user is assumed to be $R_{n}^{\min}=R_{\min}=2 {\rm \ Mbps}, \forall n \in \mathcal{N}$, and the aggregation error tolerance of the AirFL users is $\varepsilon_0 = 0.01$.
The number of reflecting elements is set as $M=20$, the quantization resolution is $b=2$, and the default weight parameter is $\lambda=0.5$, unless otherwise stated.

In order to validate the effectiveness of our proposed algorithms for the hybrid network with non-ideal phase shifts (labeled `Discrete RIS'), the following schemes are considered as benchmarks:
\begin{itemize}
	\item[\rmnum{1}.] Continuous RIS: The phase shift of each reflecting element can be tuned with arbitrary continuous value from $0$ to $2 \pi$, i.e., $b \rightarrow \infty$. In this way, the combinatorial constraint in (\ref{original_constraint_phase_shift}) can be relaxed as $\theta_{m} \in [0, 2 \pi), \forall m \in \mathcal{M}$. As a result, the subproblem of phase shifts design is directly solved by the SDR method, and the Gaussian randomization approach can be recalled if the obtained solution is not rank-one.
	\item[\rmnum{2}.] Random RIS: The phase shift of each reflecting element is initialized with random value from $0$ to $2 \pi$, and then set to discrete value in $\mathcal{A}$ via the method proposed in (\ref{RIS_problem_discrete_phase_recover}). By doing so, only the transceiver design problem (\ref{transceiver_problem}) needs to be optimized at each iteration.
	\item[\rmnum{3}.] Relaxed QoS: By setting $R_{\min}=0$, the QoS requirements requested in constraints (\ref{original_constraint_QoS}) can be ignored. In this case, the optimal transmit power at NOMA users is the full power transmission (refer to \textit{Proposition \ref{corollary_transmit_power}}), which also can be deemed as the upper bound of the communication rate that can be obtained by the NOMA users with other variables fixed.
	\item[\rmnum{4}.] Relaxed MSE: By setting $\varepsilon_0 \rightarrow \infty$, the MSE requirements requested in constraints (\ref{original_constraint_MSE}) can be ignored. In this case, the optimal receive scalar for AirFL users can be obtained in (\ref{optimal_receive_scalar}) (refer to \textit{Proposition \ref{corollary_receive_scalar}}), which also can be deemed as the upper bound of the computation rate that can be achieved by AirFL users with optimized transmit power and phase shifts.	
\end{itemize}

The convergence accuracy of the proposed algorithms is set as $10^{-6}$ for striking a balance between performance and complexity.
All presented results have been averaged over 2000 independent trials of the user positions and the channel realizations.


\subsection{Performance Evaluation}

\begin{figure} [t!]
	\centering
	\includegraphics[width=3.5 in]{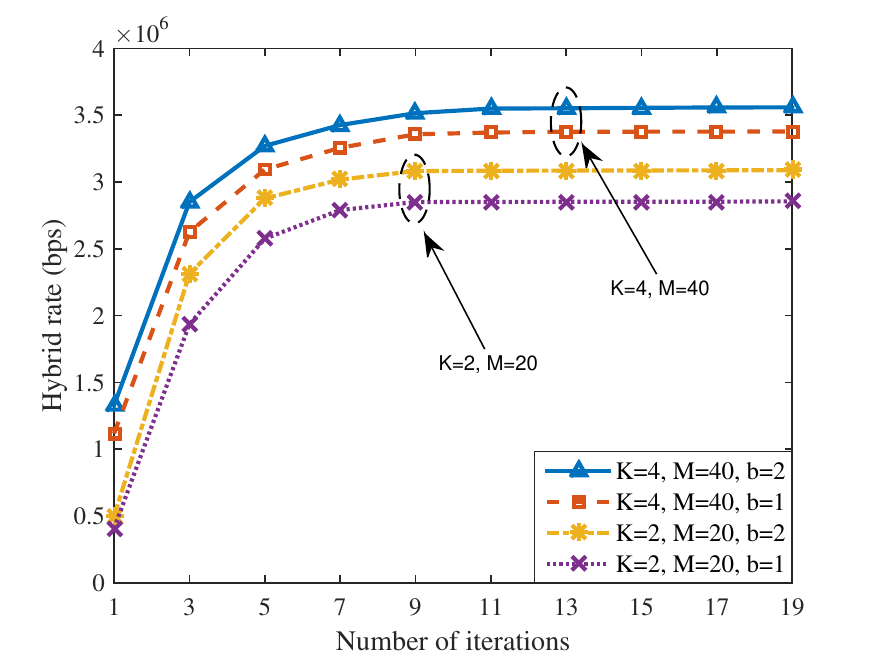}
	\caption{Hybrid rate versus the number of iterations.}
	\label{hybird_rate_vs_iteration}
\end{figure}

\subsubsection{Convergence of the proposed algorithm}
In Fig. \ref{hybird_rate_vs_iteration}, the convergence performance of the proposed algorithm versus the iteration number is depicted, which illustrates that the proposed AO-based algorithm can converge under different number of users, reflecting elements and quantization resolution.
Furthermore, it is observed that the hybrid rate grows quickly with the number of iterations, which also increases with the number of users and reflecting elements, owing to the fact that there are more variables to be optimized over iterations.
To be specific, when $K=2$ and $M=20$, the proposed algorithm needs only $9$ iterations to satisfy the convergence accuracy, while it converges with around $13$ iterations when $K=4$ and $M=40$.
From this figure, it also can be seen that with higher quantization resolution, the configuration of the RIS becomes more flexible and thus achieves better convergence performance.

\begin{figure} [t!]
	\centering
	\includegraphics[width=3.5 in]{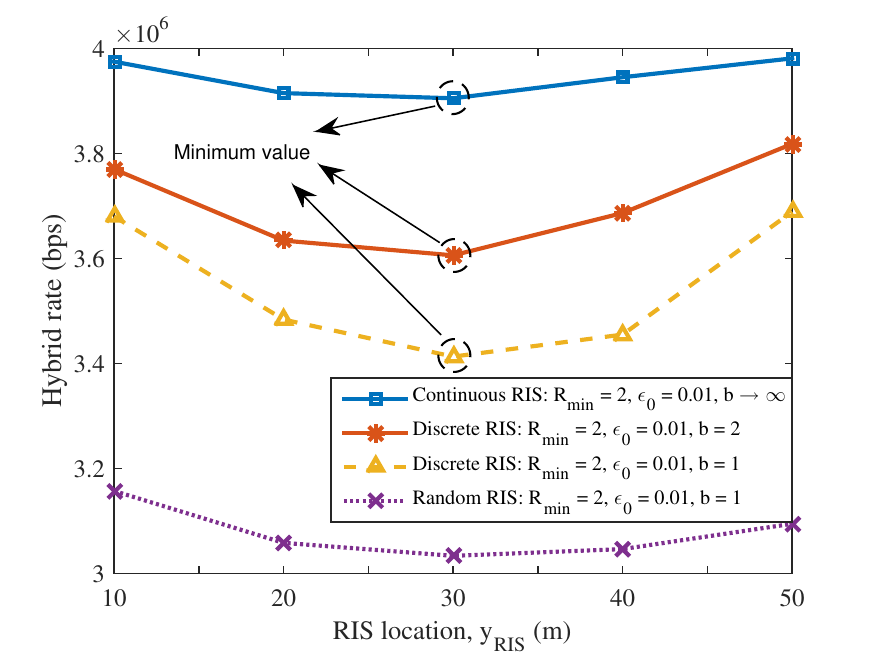}
	\caption{Hybrid rate versus the location of RIS coordinate.}
	\label{hybird_rate_vs_RIS_location}
\end{figure}

\subsubsection{Impact of the RIS location}
In Fig. \ref{hybird_rate_vs_RIS_location}, the impact of the RIS location on the achievable hybrid rate is evaluated.
In this figure, the locations of the BS and RIS are reset as $(0,0,0)$ and $(0,y_{RIS},0)$, respectively, and the users are randomly placed in a circle centered at $(0, 60, 0)$ with radius ${5 \ \rm m}$.
By moving the RIS from $y_{RIS}={10 \ \rm m}$ (BS side) to $y_{RIS}={50 \ \rm m}$ (user side), it is interesting to observe that the hybrid rate achieved by the proposed algorithm first decrease with $y_{RIS} \ ({10 \ \rm m} \le y_{RIS} \le {30 \ \rm m})$ and then increase with $y_{RIS} \ ({30 \ \rm m} \le y_{RIS} \le {50 \ \rm m})$ after achieving their minimum values at $y_{RIS}={30 \ \rm m}$ (midpoint).
This is mainly because the double fading effect experienced by the reflective link.
Similar to the analysis in \cite{Zuo2020Resource}, we focus on the large-scale path loss of the combined channel gain, i.e.,
$L_0 L_i = \varsigma_0^2 \left( d_{0} d_{i} \right) ^{-\alpha}, \forall i \in \mathcal{I}$.
Note that $d_{0} + d_{i} \simeq 60, \forall i \in \mathcal{I}$, thus the minimal value of the combined channel gain can be obtained at $d_{0} = d_{i} \simeq 30$.
Namely, it is harmful for hybrid rate maximization to deploy the RIS at the midpoint between the BS and the users.
Therefore, the system performance can be significantly improved by carefully selecting the location of the RIS, which provides guidance for the practical deployment of the RIS.

\begin{figure} [t!]
	\centering
	\includegraphics[width=3.5 in]{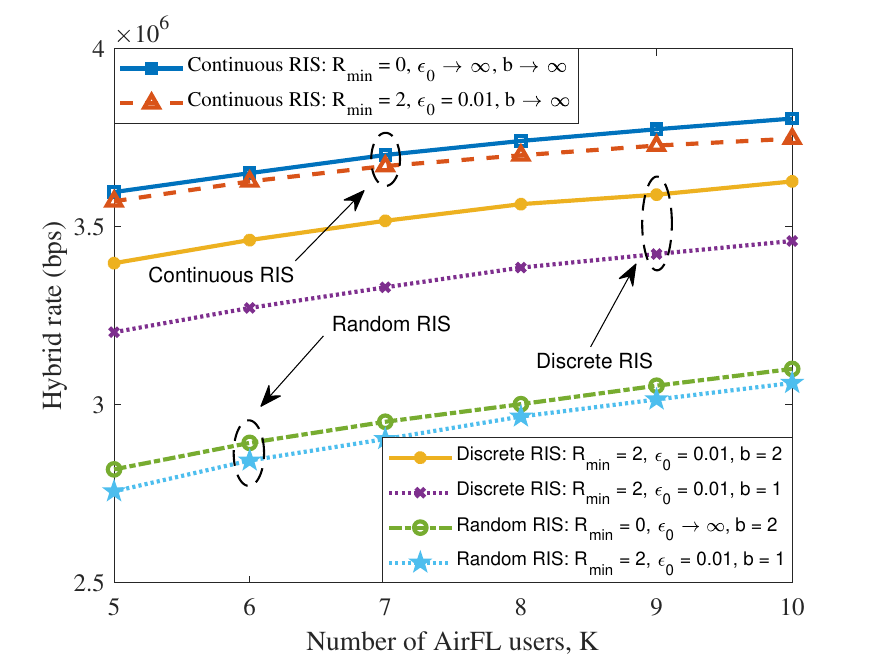}
	\caption{Hybrid rate versus the number of AirFL users.}
	\label{hybrid_rate_vs_AirFL_users}
\end{figure}

\begin{figure} [t!]
	\centering
	\includegraphics[width=3.5 in]{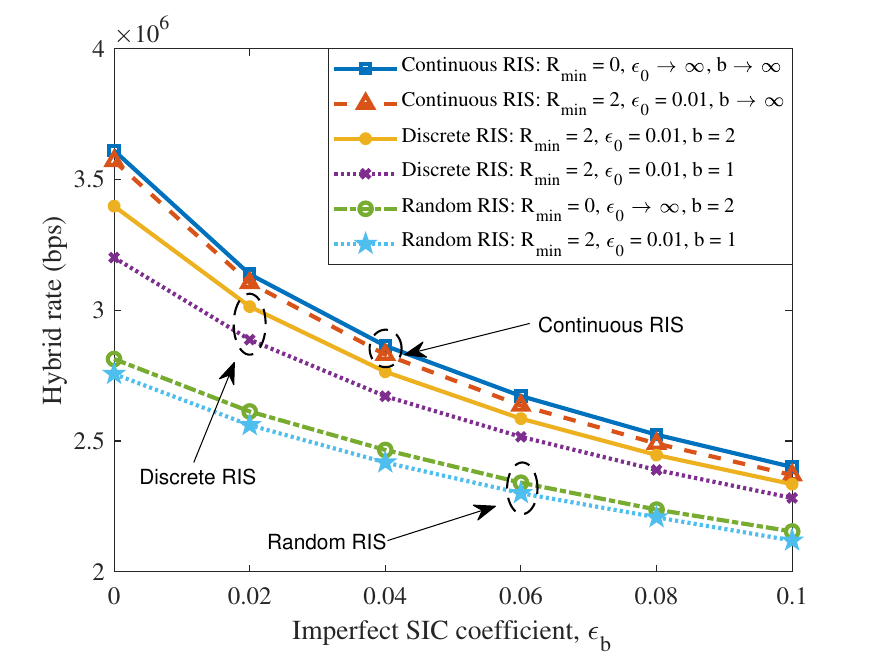}
	\caption{Hybrid rate versus the imperfect SIC coefficient.}
	\label{hybrid_rate_vs_imperfect_SIC}
\end{figure}


\subsubsection{Impact of the number of AirFL users}
In Fig. \ref{hybrid_rate_vs_AirFL_users}, we investigate the performance of the proposed RIS-aided hybrid network versus different number of AirFL users in the unified framework.
Here the number of NOMA users is fixed as we focus on the impact of AirFL users.
One can observe from Fig. \ref{hybrid_rate_vs_AirFL_users}, for all considered schemes, the achievable hybrid rate will rise steadily as the number of AirFL users grows gradually in the system.
This verifies that the remarks still hold true when the number of users increases.
Not surprisingly, the schemes of `Continuous RIS’ are capable of outperforming other benchmarks with/without QoS and MSE constraints.
This shows the great strength of high-resolution RIS in mitigating communication interference and enhancing function aggregation in the integrated network, despite this may increase the production cost of RIS.
To be more specific, when $K=8$ in the `Discrete RIS' scheme, we find that the setting of $b=2$ can achieve about $5.3\%$ higher hybrid rate than that of $b=1$, which also validates the effectiveness of the proposed relaxation-then-quantization scheme in tackling reflection design problem.

\subsubsection{Impact of the imperfect SIC}
In Fig. \ref{hybrid_rate_vs_imperfect_SIC}, we show the achievable hybrid rate under different values of the imperfect SIC coefficient.
In order to compare the different effects of increasing the SIC coefficient and the number of AirFL users on the hybrid rate, the starting points in Fig.~\ref{hybrid_rate_vs_imperfect_SIC} are the same as that in Fig. \ref{hybrid_rate_vs_AirFL_users}.
From Fig. \ref{hybrid_rate_vs_imperfect_SIC}, we observe that the achievable hybrid rate of all considered schemes decreases with the imperfect SIC coefficient $\epsilon_b$.
This matches our intuition that imperfect SIC leads to a higher residual interference than perfect SIC, thus reducing the achievable rate.
This is because when the value of $\epsilon_b$ increases, both NOMA users and AirFL users experience serious interference, which degrades the communication rate and the computation rate at the same time, resulting in the decrease of hybrid rate.

\begin{figure} [t!]
	\centering
	\includegraphics[width=3.5 in]{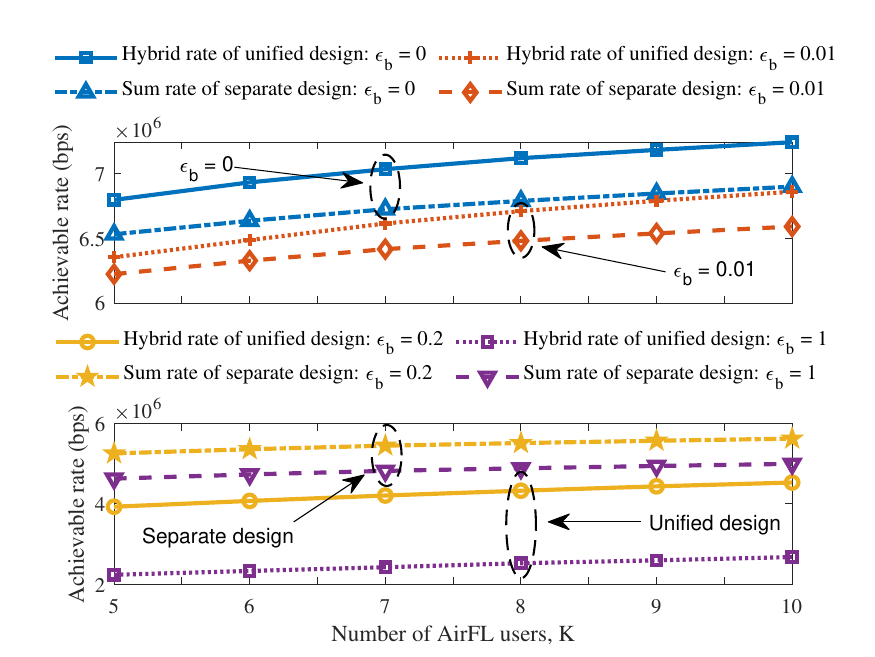}
	\caption{Achievable rate comparison of separate design and unified design with different values of the imperfect SIC coefficient and different numbers of AirFL users.} 
	\label{hybrid_and_sparate_rate_vs_AirFL_users}
\end{figure}

\begin{figure} [t!]
	\centering
	\includegraphics[width=3.5 in]{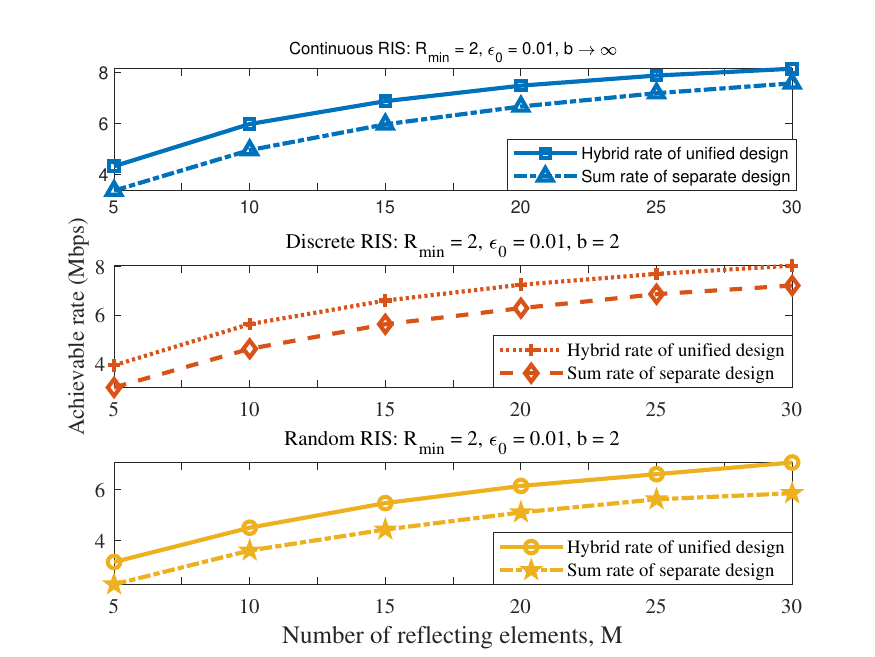}
	\caption{Achievable rate comparison of separate design and unified design versus the number of reflecting elements under different RIS settings.} 
	\label{hybrid_and_sparate_rate_vs_M}
\end{figure}


\subsubsection{Separate design versus unified design}
In Fig. \ref{hybrid_and_sparate_rate_vs_AirFL_users}, we compare the achievable rate of separate design and unified design with different values of the imperfect SIC coefficient and different numbers of AirFL users.
For a fair comparison of two diametrically opposite paradigms, both designs occupy the same time slot length and utilize the whole system bandwidth.
In the separate design, two types of users are alternatively scheduled and the associated rates are independently maximized.
To be specific, half of the time slot is only used to serve NOMA users for the purpose of $\max \ R_{\rm NOMA}$, while the remaining AirFL users are served by the rest time slot with the aim of $\max \ R_{\rm AirFL}$.
In contrast, all hybrid users in the unified design are simultaneously served in a non-orthogonal manner with the objective of $\max \ R_{\rm NOMA} + R_{\rm AirFL}$.
One can notice that the achievable rates of both separate and unified designs grow with the number of AirFL users.
Specifically, for $\epsilon_b=0$ (perfect SIC) and $\epsilon_b=0.01$ (nearly perfect SIC), we see that the hybrid rate of unified design is always higher than the sum rate of separate design.
However, if we increase $\epsilon_b$ to a larger value, an interesting point is that the hybrid rate of unified design will be lower than the sum rate of separate design, e.g., $\epsilon_b=0.2$ (imperfect SIC) and $\epsilon_b=1$ (no SIC).
This is because both NOMA and AirFL users in the unified design will experience more severe inter-user interference as the increase of $\epsilon_b$.
Whereas, AirFL users in the separate design do not experience any interference from NOMA users due to time division.
This also validates the importance of efficient interference management for the proposed unified design.
In Fig. \ref{hybrid_and_sparate_rate_vs_M}, considering $K=10$ AirFL users with perfect SIC, we compare the achievable rate of separate design and unified design versus the number of reflecting elements.
It can be found that the proposed unified design outperforms separate design under different RIS settings.
Specifically, for the discrete RIS with $M=20$, the hybrid rate of unified design has a $15.4\%$ increment compared to the sum rate of separate design.

\begin{figure*}[t]
	\centering
	\subfloat[Communication rate versus M]{%
		\includegraphics[width=.32 \textwidth]{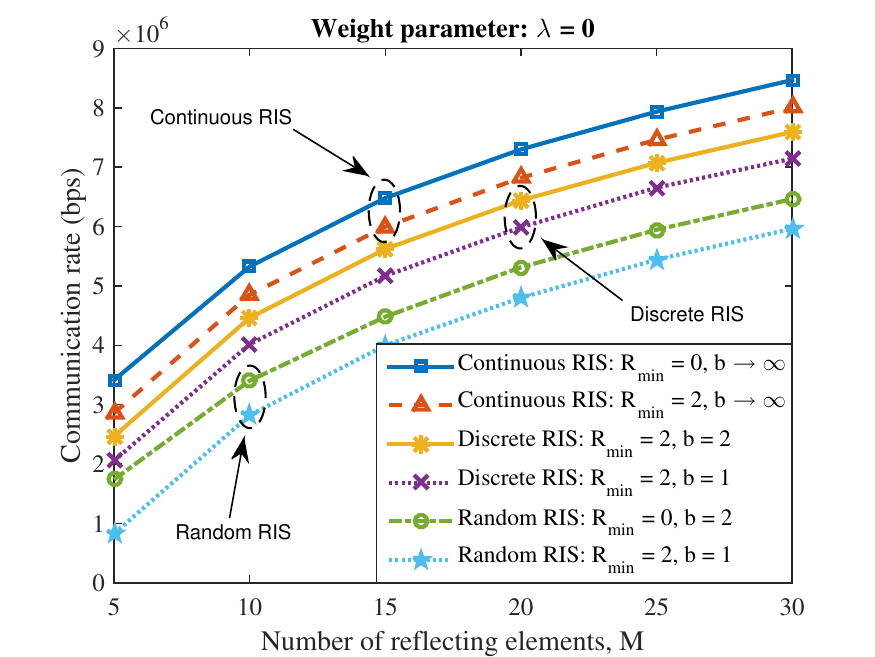} \label{NOMA_rate_vs_M} }\hfill
	\subfloat[Hybrid rate versus M]{%
		\includegraphics[width=.32 \textwidth]{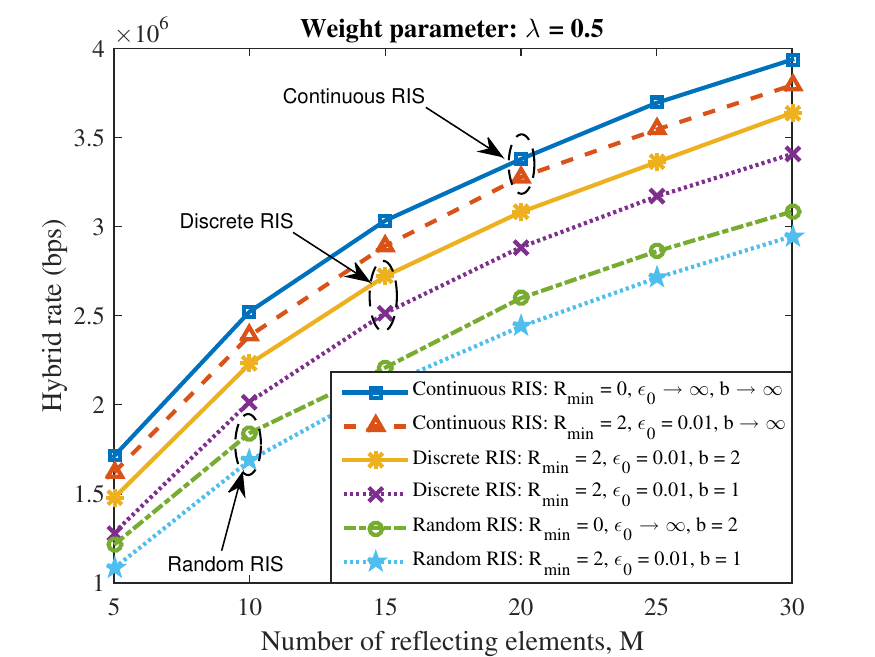} \label{Hybrid_rate_vs_M} } \hfill
	\subfloat[Computation rate versus M]{%
		\includegraphics[width=.32 \textwidth]{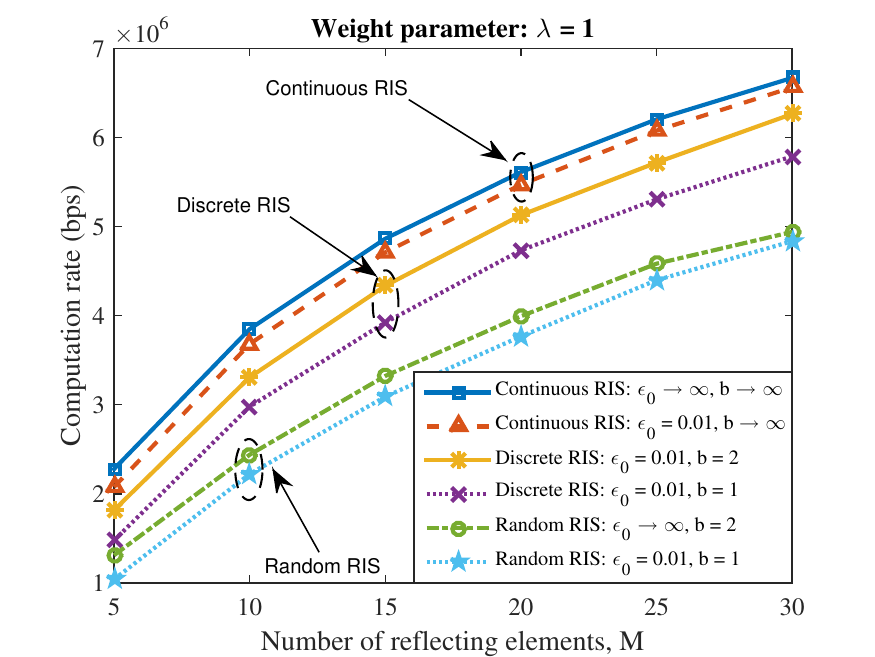} \label{AirFL_rate_vs_M} } \\
	\subfloat[Communication rate versus $P_{\max}$]{%
		\includegraphics[width=.32 \textwidth]{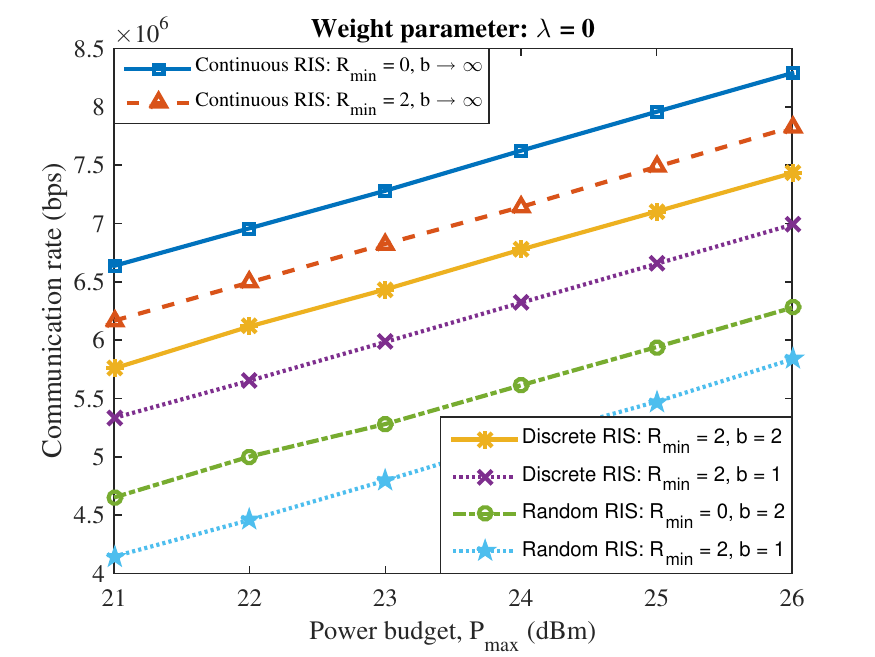} \label{NOMA_rate_vs_P_max} } \hfill
	\subfloat[Hybrid rate versus $P_{\max}$]{%
		\includegraphics[width=.32 \textwidth]{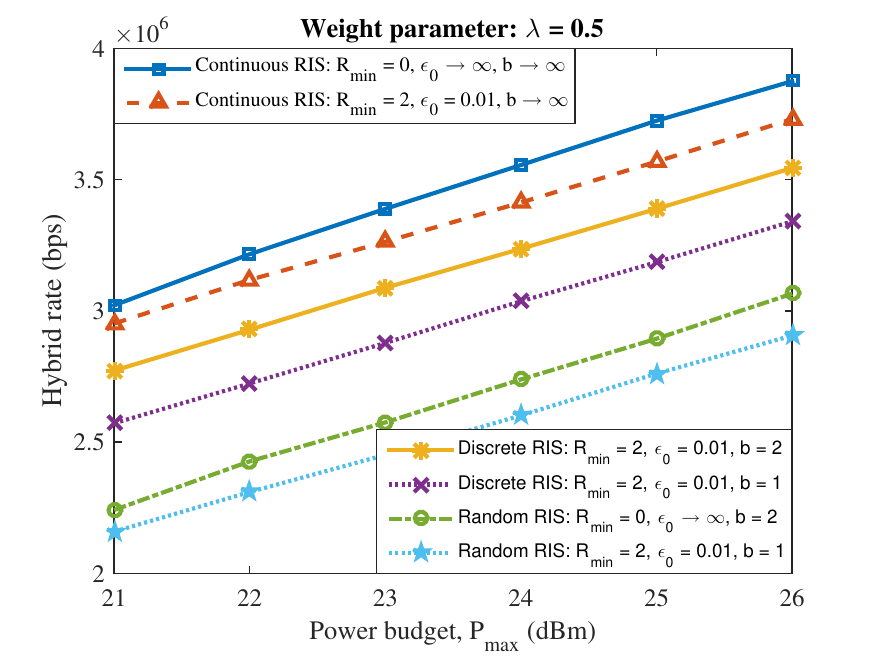} \label{Hybrid_rate_vs_P_max} } \hfill
	\subfloat[Computation rate versus $P_{\max}$]{%
		\includegraphics[width=.32 \textwidth]{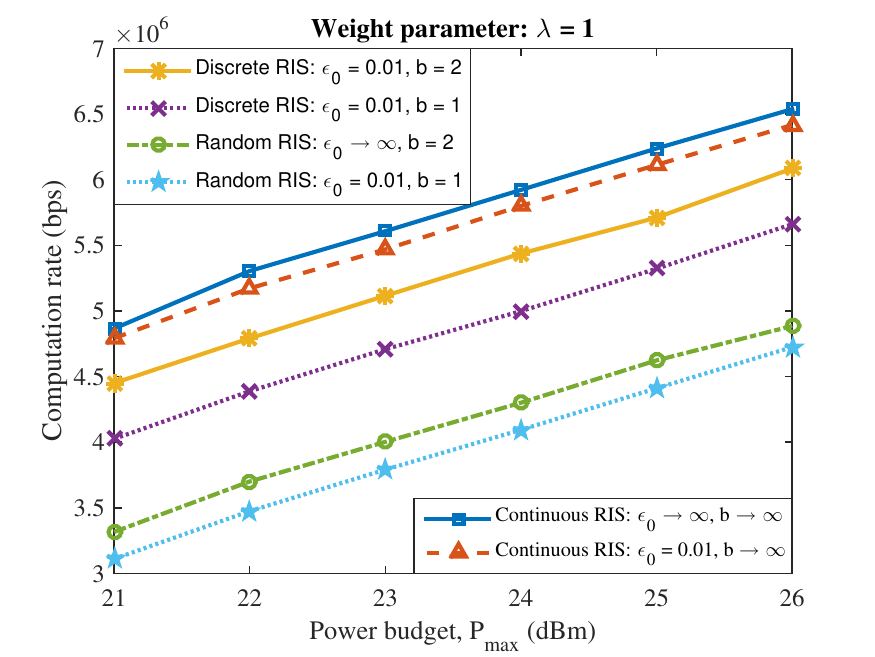} \label{AirFL_rate_vs_P_max} }
	\caption{Achievable rate under different settings: communication rate ($\lambda = 0$), hybrid rate ($\lambda = 0.5$) and computation rate ($\lambda = 1$) versus the number of reflecting elements ($M \in [5 \ 30]$) and the power budget ($P_{\max} \in [21 \ 26]$).}\label{Achievable_rate}
\end{figure*}

\subsubsection{Impact of the number of reflecting elements}
In Fig. \ref{Achievable_rate}, the hybrid rate performance versus the number of reflecting elements at the RIS and the power budget at the users are illustrated.
For comparison, the following three cases are considered, i.e.,
\rmnum{1}) $\lambda = 0$, i.e., there are only $2$ NOMA users;
\rmnum{2}) $\lambda = 0.5$, i.e., there are $4$ hybrid users;
\rmnum{3}) $\lambda = 1$, i.e., there are only $2$ AirFL users.
Specifically, the first row of Fig. \ref{Achievable_rate} evaluates the impact of the number of reflecting elements on the achievable rate under three cases.
On the whole, the following two insights can easily be drawn.
The first one is that the achievable rate under all cases grows with the number of reflecting elements $M$, which reveals the effectiveness of having more reflecting elements at the RIS.
Besides, it is observed that for hybrid rate of $3 \times 10^6 {\rm \ bps}$, we need to deploy around 20 reflecting elements with $b=2$.
In contrast, with the same hybrid rate performance, we can alternatively use about 25 reflecting elements with $b=1$ at the RIS.
Therefore, another interesting insight is that the RIS-aided hybrid network provides more flexibility to balance the trade-off between the number of passive elements and their quantization resolution.

\subsubsection{Impact of the power budget}
The second row of Fig. \ref{Achievable_rate} plots the achievable rate versus the maximum transmit power at users.
Similar insights mentioned above are omitted here for brevity.
Distinctively, one can notice that the achievable rate of all considered schemes grows linearly with $P_{\max}$, which is an attractive trend.
But the available power is limited in practice, especially for these low-cost devices.
Additionally, the achievable rate can be significantly improved by jointly optimizing the transceiver and the phase shifts at the RIS, regardless of the values of $\lambda$ and $P_{\max}$.
More particularly, when $P_{\max}=25$ dBm and $b=1$, it can be obtained that the RIS-aided wireless networks are capable of providing up to $22.3\%$, $16.9\%$ and $20.6\%$ higher achievable rate than the random RIS schemes under three cases (i.e., $\lambda = 0, 0.5, 1$).
This is because that the RIS is beneficial to enhance the channel quality for NOMA users and reduce the aggregation error for AirFL users by carefully adjusting the decoding order of all users.



\section{Conclusion} \label{section_conclusion}
In this paper, we proposed a novel RIS-aided hybrid network to effectively integrate AirFL and NOMA into an elastic framework.
The role of RIS is twofold: \rmnum{1}) efficiently enhance the channel condition and networking coverage that may be unfavorable due to complete blockage, and \rmnum{2}) flexibly change the channel gains of hybrid users to achieve an on-demand decoding order that is efficient for NOMA, for AirFL, and for the combination of both.
Specifically, we investigated an intractable resource allocation problem of the RIS-aided hybrid network by jointly optimizing the transmit power, the receive scalar, and the phase shifts for hybrid rate maximization while meeting QoS requirements as well as MSE tolerance.
To address this challenging mixed-combinatorial optimization problem, we devised an alternating optimization algorithm to tackle the decoupled non-convex subproblems by iteratively adopting approximation techniques such as DC programming, SCA, and SDR methods.
Besides, regarding the reduced subproblems of transmit power allocation and receive scalar design, the optimal solutions were derived in closed form.
Furthermore, the complexity and convergence performance were analyzed.
Finally, simulation results demonstrated that the proposed RIS-aided network can simultaneously serve hybrid users on the entire bandwidth effectively, and the achievable hybrid rate can be significantly enhanced by tuning the RIS judiciously.

\appendices
\section{Proof of Lemma \ref{lemma_1}} \label{proof_of_lemma_1}
According to (\ref{hat_h}), with given $\alpha=2$, the combined channel gain can be expanded as
\begin{align}
\left| \bar{h}_{i} \right| ^2 = \boldsymbol{v}^{\rm H} \widetilde{ \boldsymbol{\Phi} }_{i} \widetilde{ \boldsymbol{\Phi} }_{i}^{\rm H} \boldsymbol{v} = \boldsymbol{v}^{\rm H} \boldsymbol{\Lambda}_{i} \boldsymbol{v}, \ \forall i \in \mathcal{I}.
\end{align}
Note that $\boldsymbol{v}^{\rm H} \boldsymbol{\Lambda}_{i} \boldsymbol{v} = {\rm tr} \left( \boldsymbol{\Lambda}_{i} \boldsymbol{v} \boldsymbol{v}^{\rm H} \right), \ \forall i \in \mathcal{I} $, we thus have
\begin{align}
\left| \bar{h}_{i} \right| ^2 = {\rm tr} \left( \boldsymbol{\Lambda}_{i} \boldsymbol{v} \boldsymbol{v}^{\rm H} \right)
= {\rm tr} \left( \boldsymbol{\Lambda}_{i} \widetilde{ \boldsymbol{V} } \right), \ \forall i \in \mathcal{I},
\end{align}
where $\boldsymbol{\Lambda}_{i} = \widetilde{ \boldsymbol{\Phi} }_{i} \widetilde{ \boldsymbol{\Phi} }_{i}^{ \rm H }$, $ \widetilde{ \boldsymbol{V} } = \boldsymbol{v} \boldsymbol{v}^{\rm H}$ and $\widetilde{ \boldsymbol{V} } \succeq \boldsymbol{0}$.
The proof is completed.

\section{Proof of Lemma \ref{lemma_2}} \label{proof_of_lemma_2}
When $\widehat{ \boldsymbol{\Phi}}_{k} = \frac{ a \varsigma_0 \boldsymbol{\Phi}_{k} p_{k} }{ d_{0} d_{k} }, \forall k \in \mathcal{K}$, the left-hand-side (LHS) in (\ref{lemma_trace_tilde_V}) can be expanded as
\begin{align}
\left| a \bar{h}_{k} p_{k} \right| ^{2} = \left| \boldsymbol{v}^{\rm H} \widehat{ \boldsymbol{\Phi}}_{k} \right| ^{2}
= \boldsymbol{v}^{\rm H} \widehat{ \boldsymbol{\Phi}}_{k} \widehat{ \boldsymbol{\Phi}}_{k}^{\rm H} \boldsymbol{v}, \ \forall k \in \mathcal{K}.
\end{align}
Since $\widetilde{ \boldsymbol{\Lambda} }_{k} = \widehat{ \boldsymbol{\Phi}}_{k} \widehat{ \boldsymbol{\Phi}}_{k}^{\rm H}, \forall k \in \mathcal{K}$, then we can obtain
\begin{align}
\left| a \bar{h}_{k} p_{k} \right| ^{2}
= {\rm tr} \left( \widetilde{ \boldsymbol{\Lambda} }_{k} \boldsymbol{v} \boldsymbol{v}^{\rm H} \right)
= {\rm tr} \left( \widetilde{ \boldsymbol{\Lambda} }_{k} \widetilde{ \boldsymbol{V} } \right), \ \forall k \in \mathcal{K},
\end{align}
which completes the proof.

\section{Proof of Lemma \ref{lemma_3}} \label{proof_of_lemma_3}
Similarly, the LHS in (\ref{lemma_trace_hat_V}) can be rewritten as
\begin{equation}
\left| a \bar{h}_{k} p_{k} - 1 \right| ^{2} = \left| \boldsymbol{v}^{\rm H} \widehat{ \boldsymbol{\Phi}}_{k} - 1 \right| ^{2}, \ \forall k \in \mathcal{K}.
\end{equation}
Then, it can be expanded as 
\begin{align}
\left| a \bar{h}_{k} p_{k} - 1 \right| ^{2}
& = \boldsymbol{v}^{\rm H} \widehat{ \boldsymbol{\Phi}}_{k} \widehat{ \boldsymbol{\Phi}}_{k}^{\rm H} \boldsymbol{v} - \boldsymbol{v}^{\rm H} \widehat{ \boldsymbol{\Phi}}_{k} - \widehat{ \boldsymbol{\Phi}}_{k}^{\rm H} \boldsymbol{v} + 1 \nonumber \\
& = \boldsymbol{\bar{v}}^{\rm H} \widehat{ \boldsymbol{\Lambda}}_{k} \boldsymbol{\bar{v}} + 1, \ \forall k \in \mathcal{K},
\end{align}
where
\begin{equation}
\widehat{ \boldsymbol{\Lambda}}_{k} = \left[ \begin{array}{cc}
\widehat{ \boldsymbol{\Phi}}_{k} \widehat{ \boldsymbol{\Phi}}_{k}^{\rm H} & - \widehat{ \boldsymbol{\Phi}}_{k} \\
- \widehat{ \boldsymbol{\Phi}}_{k}^{\rm H} & 0 
\end{array} 
\right ]
\text{and} \
\boldsymbol{\bar{v}} = \left[ \begin{array}{c}
\boldsymbol{v} \\
1
\end{array} 
\right ].
\end{equation}
Note that $\boldsymbol{\bar{v}}^{\rm H} _{k} \widehat{ \boldsymbol{\Lambda}}_{k} \boldsymbol{\bar{v}} = {\rm tr} \left( \widehat{ \boldsymbol{\Lambda}}_{k} \boldsymbol{\bar{v}} \boldsymbol{\bar{v}}^{\rm H} \right), \ \forall k \in \mathcal{K} $, we thus have
\begin{equation}
\left| a \bar{h}_{k} p_{k} - 1 \right| ^{2}
= {\rm tr} \left( \widehat{ \boldsymbol{\Lambda}}_{k} \boldsymbol{\bar{v}} \boldsymbol{\bar{v}}^{\rm H} \right) + 1
= {\rm tr} \left( \widehat{ \boldsymbol{\Lambda}}_{k} \boldsymbol{V} \right) + 1, \ \forall k \in \mathcal{K},
\end{equation}
where $\boldsymbol{V} = \boldsymbol{\bar{v}} \boldsymbol{\bar{v}}^{\rm H}$, while $\boldsymbol{V} \succeq \boldsymbol{0}$ and ${\rm rank} ( \boldsymbol{V} ) = 1$.
This completes the proof.

\bibliographystyle{IEEEtran}
\bibliography{IEEEabrv,ref}

\end{document}